\documentclass[superscriptaddress,aps,preprintnumbers,amsmath,amssymb,prd,nofootinbib,preprint,longbibliography]{revtex4-1}
\pdfoutput=1

\usepackage{graphicx,array}
\usepackage{color}
\usepackage{latexsym}
\usepackage{amsthm}
\usepackage{amsmath}
\usepackage{amssymb}
\usepackage{hyperref} 
\usepackage{bbold}
\usepackage{mathtools}
\usepackage{enumitem}

\numberwithin{equation}{section}

\newcommand{\itemtitled}[1]{\item {\small \emph{#1}}\\}

\makeatletter
\renewcommand{\p@subsection}{}
\renewcommand{\p@subsubsection}{}
\makeatother

\def\simgt{\mathrel{\lower2.5pt\vbox{\lineskip=0pt\baselineskip=0pt
           \hbox{$>$}\hbox{$\sim$}}}}
\def\simlt{\mathrel{\lower2.5pt\vbox{\lineskip=0pt\baselineskip=0pt
           \hbox{$<$}\hbox{$\sim$}}}}

\newcommand{\be}{\begin{equation}}
\newcommand{\ee}{\end{equation}}
\newcommand{\bea}{\begin{eqnarray}}
\newcommand{\eea}{\end{eqnarray}}

\newcommand{\eV}{\textrm{ eV}}

\newcommand{\GeV}{\textrm{ GeV}}
\newcommand{\TeV}{\textrm{ TeV}}
\newcommand{\gsim}{\lower.7ex\hbox{$\;\stackrel{\textstyle>}{\sim}\;$}}
\newcommand{\lsim}{\lower.7ex\hbox{$\;\stackrel{\textstyle<}{\sim}\;$}}

\newcommand{\MPl}{M_{\rm Pl}}

\definecolor{nicered}{rgb}{0.7,0.1,0.1}
\definecolor{nicegreen}{rgb}{0.1,0.5,0.1}
\hypersetup{colorlinks,citecolor= blue,linkcolor= black}

\begin{document}

\title{ \textit{\textbf{R}}-Parity Violation Axiogenesis}

\author{Raymond T. Co}
\affiliation{William I. Fine Theoretical Physics Institute, School of Physics and Astronomy, University of Minnesota, Minneapolis, MN 55455, USA}
\author{Keisuke Harigaya}
\affiliation{Theoretical Physics Department, CERN, Geneva, Switzerland}
\affiliation{School of Natural Sciences, Institute for Advanced Study, Princeton, NJ 08540, USA}
\author{Zachary Johnson}
\affiliation{Leinweber Center for Theoretical Physics, Department of Physics, University of Michigan, Ann Arbor, MI 48109, USA}
\author{Aaron Pierce}
\affiliation{Leinweber Center for Theoretical Physics, Department of Physics, University of Michigan, Ann Arbor, MI 48109, USA}

\date{\today}

\begin{abstract}
We show that the rotation of the QCD axion field, aided by $B-L$ violation from supersymmetric $R$-parity violating couplings, can yield the observed baryon abundance.  Strong sphaleron processes transfer the angular momentum of the axion field into a quark chiral asymmetry, which $R$-parity violating couplings convert to the baryon asymmetry of the Universe. We focus on the case of dimensionless $R$-parity violating couplings with textures motivated by grand unified theories and comment on more general scenarios. The axion decay constant and mass spectrum of supersymmetric particles are constrained by Big Bang nucleosynthesis, proton decay from the $R$-parity violation, and successful thermalization of the Peccei-Quinn symmetry breaking field. Axion dark matter may be produced by the axion rotation via the kinetic misalignment mechanism for axion decay constants below $10^{10}$ GeV, or by the conventional misalignment mechanism for  $10^{11\mathchar`-12}$ GeV. The viable parameter region can be probed by proton decay and axion searches. This scenario may also have connections with collider experiments, including searches for long-lived particles, and observations of gravitational waves. 
\end{abstract}

\preprint{UMN-TH-4104/21, FTPI-MINN-21-21, CERN-TH-2021-147, LCTP-21-25}

\maketitle

\vspace{1cm}

\maketitle

\begin{quote} \small

\end{quote}

\tableofcontents

\newpage

\section{Introduction}
\label{sec:intro}

The strong CP problem can elegantly be solved via the introduction of a Peccei-Quinn (PQ) symmetry~\cite{Peccei:1977hh, Peccei:1977ur} with its accompanying axion~\cite{Weinberg:1977ma,Wilczek:1977pj}.  The axion also leads to interesting cosmological possibilities.  For example, the axion is an attractive cold dark matter candidate via the misalignment mechanism~\cite{Preskill:1982cy, Dine:1982ah,Abbott:1982af}, where the axion field begins its evolution at rest but displaced from the minimum of its potential.

Recently, the possibility that the axion receives a ``kick"  and rotates in field space has been explored, with implications for both dark matter and baryogenesis.  The kick may arise from explicit PQ symmetry breaking by higher dimensional operators.  The effects of these operators are enhanced because of a large initial field value of the radial direction of the PQ symmetry breaking field, as in the Affleck-Dine mechanism~\cite{Affleck:1984fy,Dine:1995kz}. The rotation of the axion corresponds to a non-zero PQ charge, which is approximately conserved once the radial direction moves to lower field values and the higher dimensional operators become suppressed. 
This charge is transferred into a quark chiral asymmetry via the strong sphaleron processes~\cite{Co:2019wyp} due to the quantum anomaly of the PQ symmetry with respect to the strong interactions.
The quark chiral asymmetry may be further transferred into a baryon-antibaryon asymmetry via baryon number-violating processes. This mechanism has been dubbed ``axiogenesis"~\cite{Co:2019wyp}.
In the minimal case, the quark chiral asymmetry may be transferred into the baryon-antibaryon asymmetry via electroweak sphalerons~\cite{Klinkhamer:1984di,Kuzmin:1985mm}.

\begin{figure}
    \centering
\includegraphics[width=\linewidth]{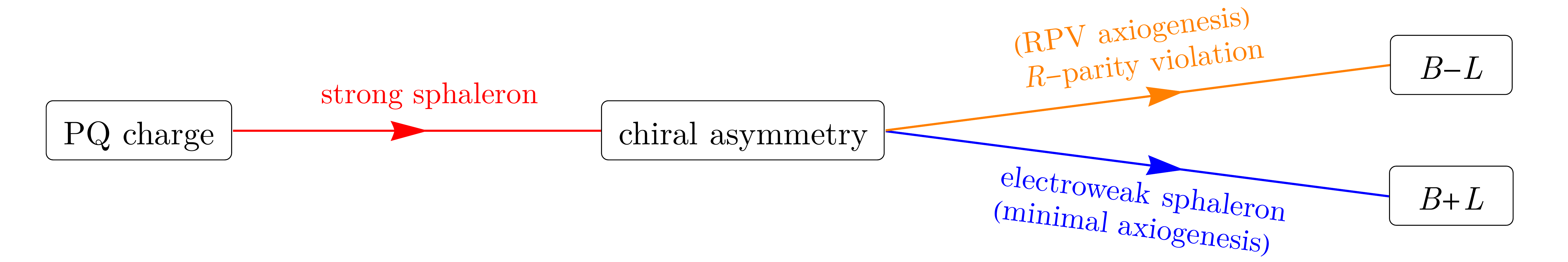}
    \caption{Transfer of asymmetries for minimal axiogenesis and RPV axiogenesis}
    \label{fig:schematic_RPV}
\end{figure}

A schematic of the charge transfer in axiogenesis is shown in Fig.~\ref{fig:schematic_RPV}. In the minimal scenario, only electroweak sphaleron processes provide baryon number violation, and only $B+L$ rather than $B-L$ is produced.  Any baryon asymmetry produced prior to the electroweak phase transition is washed out by electroweak sphaleron processes, and the final baryon asymmetry is fixed at the electroweak phase transition. The kinetic energy of the rotation may also transform into the axion dark matter density, which is called the kinetic misalignment mechanism (KMM)~\cite{Co:2019jts}. 
However, after requiring that the KMM should not overproduce axion dark matter and assuming the standard electroweak phase transition temperature, the baryon asymmetry produced by minimal axiogenesis is too small.
To explain the observed baryon asymmetry, new physics beyond the Standard Model (SM) and the QCD axion is therefore required; see Refs.~\cite{Co:2019wyp,Co:2020xlh,Co:2020jtv,Harigaya:2021txz,Chakraborty:2021fkp}. 
Requiring the correct baryon abundance without overproducing axion dark matter imposes constraints on this new physics.

In this paper, we consider supersymmetry,  one of the best-motivated frameworks beyond the SM.
Supersymmetry is the unique extension of the spacetime symmetry~\cite{Haag:1974qh}, leads to precise gauge coupling unification~\cite{Dimopoulos:1981yj,Dimopoulos:1981zb,Sakai:1981gr,Ibanez:1981yh,Einhorn:1981sx,Marciano:1981un}, and significantly relaxes the electroweak hierarchy problem~\cite{Maiani:1979cx,Veltman:1980mj,Dimopoulos:1981zb, Witten:1981nf,Kaul:1981wp}. In the QCD axion context, supersymmetry can also stabilize the hierarchy between the PQ symmetry breaking scale and the Planck scale against quantum corrections even if the PQ symmetry breaking field is a fundamental scalar. In supersymmetric theories without $R$-parity, $B-L$ is violated at the renormalizable level~\cite{Weinberg:1981wj}. 
The $R$-parity violation (RPV, see~\cite{Barbier:2004ez} for a review) can generate a $B-L$ asymmetry from the chiral asymmetry produced by the axion rotation, as shown in Fig.~\ref{fig:schematic_RPV}. We call this scenario ``RPV axiogenesis." Unlike the $B+L$ asymmetry of the minimal axiogenesis model, the $B-L$ asymmetry is impervious to electroweak-sphaleron washout, and so it may be produced prior to the electroweak phase transition. (Effective) $B-L$ violation from new physics is utilized also in the proposals in~\cite{Co:2020jtv,Harigaya:2021txz,Chakraborty:2021fkp}. The resultant baryon asymmetry depends on the magnitude of RPV and the masses of supersymmetric particles, rather than the electroweak phase transition temperature.

The strength of RPV impacts not just the size of the $B-L$ asymmetry, but the way in which it is produced. Borrowing terminology from dark matter production, production may occur either in the ``freeze-out" or ``freeze-in" regime. If RPV is large enough, the RPV interaction can be in thermal equilibrium in the early universe; it decouples as the temperature of the universe drops sufficiently below the masses of the superpartners, and the $B-L$ asymmetry freezes out at this temperature. On the other hand, if RPV is small, it never attains thermal equilibrium, and $B-L$ asymmetry freezes in. The freeze-in temperature at which $B-L$ is dominantly produced depends on the evolution of the axion field's angular velocity and the masses of supersymmetric particles.

In this paper, we focus on dimensionless RPV in the superpotential with an $SU(5)$ grand unified relation imposed.  This predicts nucleon decay from RPV. We find that reconciling the predicted magnitude of RPV (from successful baryogenesis) with proton decay constraints requires the sfermion mass to be above $\mathcal{O}(10-100)$ TeV for $f_a = \mathcal{O}(10^{8\mathchar`-11})$ GeV.
Future observations of proton decay, in particular $p\rightarrow K^0 \mu^+/ K^+ \bar{\nu}$, will further probe the viable parameter space. The sfermion mass is also bounded from above by requiring that the lightest supersymmetric particle (LSP) decay without disturbing Big-Bang Nucleosynthesis (BBN).  For a TeV-scale electroweakino LSP, this upper bound is intriguingly close to the lower bound from proton decay.  It is possible that long-lived LSP decays might be probed at collider experiments, such as MATHUSLA~\cite{Lubatti:2019vkf}.

The required range of the scalar mass is in remarkable agreement with scenarios without supersymmetry breaking singlet fields in the hidden sector, which we call the ``without-singlet" scenario~\cite{Giudice:1998xp,Wells:2003tf,ArkaniHamed:2004fb,Giudice:2004tc,Wells:2004di,Ibe:2006de,Acharya:2007rc,Hall:2011jd,Ibe:2011aa,Arvanitaki:2012ps,ArkaniHamed:2012gw}. (In the literature, this scenario is also sometimes referred to as mini-split SUSY, spread SUSY, pure gravity mediation, or simply the heavy scalar scenario.) Indeed, many simple dynamical supersymmetry breaking models do not contain singlet supersymmetry breaking fields. In this case, the scalar superpartners obtain soft masses at the tree-level and are heavy, while gauginos are massless at tree-level but obtain masses by one-loop quantum corrections~\cite{Randall:1998uk,Giudice:1998xp}, and they may be accessible at the LHC or near future colliders.
In this scenario, the observed Higgs mass is easily explained by quantum corrections from a large scalar top mass~\cite{Okada:1990gg,Okada:1990vk,Ellis:1990nz,Haber:1990aw}, the Polonyi~\cite{Coughlan:1983ci} and gravitino~\cite{Pagels:1981ke,Weinberg:1982zq,Khlopov:1984pf,Kawasaki:2008qe} problems are absent, and the flavor problem~\cite{Dimopoulos:1981zb,Ellis:1981ts} is significantly relaxed.

We also study the constraints on the model parameters by the successful thermalization of the rotation of the PQ symmetry breaking field. Just after the initial kick, the rotation is not completely circular and contains both angular and radial motion. The radial motion should be dissipated early enough so that the entropy production from thermalization is not too large.  Too much entropy production may unacceptably dilute the baryon asymmetry.

Unless the LSP is a light gravitino or axino, RPV forces it to decay on cosmological time scales, which means it can no longer be the dark matter. However, in our scenario, the QCD axion provides an excellent dark matter candidate.  Axion dark matter may be produced by either the kinetic~\cite{Co:2019jts} or conventional~\cite{Preskill:1982cy, Dine:1982ah,Abbott:1982af} misalignment mechanisms in the parameter space consistent with the above constraints. We also comment on the possibility of other production mechanisms.

In the majority of the allowed parameter space, the kinetic energy of the axion rotation dominates the energy density of the universe, giving rise to an axion kination era~\cite{Co:2019wyp}. A kination era is known to modify the spectrum of possible primordial gravitational waves, e.g., those produced by inflation~\cite{Giovannini:1998bp,Giovannini:1999bh,Giovannini:1999qj,Riazuelo:2000fc,Sahni:2001qp,Tashiro:2003qp,Boyle:2007zx} or local cosmic strings~\cite{Cui:2017ufi, Cui:2018rwi, Auclair:2019wcv}, while such an axion kination era imprints a unique feature on the spectrum~\cite{Co:2021lkc,Gouttenoire:2021wzu}.

A number of models of baryogenesis have been constructed using $R$-parity violating supersymmetry.  An early example uses out-of-equilibrium decays of squarks~\cite{Dimopoulos:1987rk}; see also~\cite{Claudson:1983js}.  Others utilize the decays of gauginos in a visible~\cite{Cui:2012jh,Cui:2013bta,Arcadi:2015ffa,Rompineve:2013grm} or hidden~\cite{Pierce:2019ozl} sector.  In the case of visible gauginos, often a non-canonical supersymmetry spectrum is required.  In these scenarios the generated asymmetry is proportional to the amount of RPV, and after accounting for loop factors, relatively large RPV couplings are required. Ref.~\cite{Higaki:2014eda} considers the initiation of the rotation of the Affleck-Dine field by RPV.

The idea that the baryon asymmetry could be created via the combination of baryon number violation and a non-zero velocity of a pseudo-Nambu-Goldstone boson (pNGB) field that couples to the baryon number current is considered in~\cite{Cohen:1987vi,Cohen:1988kt}, with the non-zero velocity understood as a background effective chemical potential. Ref.~\cite{Kusenko:2014uta} instead considers lepton number violation provided by Majorana neutrino masses and a pNGB that couples to the weak gauge boson. In both of these setups, the velocity of the pNGB field at the time of $B-L$ creation is driven by vacuum potential of the field.  To have sufficient velocity, the mass of the pNGB field must then be much larger than that of the QCD axion to explain the observed baryon asymmetry. In our setup, the axion velocity is induced by the kick from higher dimensional explicit PQ breaking and the resultant inertial motion. The vacuum axion mass does not play a role in baryogenesis, and it may be as small as that of the QCD axion.
Baryogenesis induced from the rotation of the PQ symmetry breaking field initiated by the same way as ours is considered in~\cite{Chiba:2003vp,Takahashi:2003db}, but those references require an interaction that violates both PQ and $B-L$ to be in thermal equilibrium.
In our setup, the QCD axion need not couple to the baryon current nor the weak gauge boson, nor does it require an interaction with a special property; an ordinary QCD axion (with its attendant coupling to the gluon) 
and supersymmetry with RPV are sufficient.

In Sec.~\ref{sec:rotating}, we review the dynamics of the rotating axion, its relationship to the generation of dark matter (via kinetic misalignment) and the baryon asymmetry (via axiogenesis), and the thermalization of the PQ symmetry breaking field. We also discuss the potential washout of the PQ charge (which could modify the baryon asymmetry we wish to generate). Here we focus on new issues that arise in the implementation of axiogenesis in supersymmetric theories.
In Sec.~\ref{sec:RPV}, after deriving the constraints from proton and LSP decay, we discuss how the mechanism proceeds in the presence of $R$-parity violating couplings in the superpotential. Appendices contain detailed discussions of the calculation of the freeze-in of baryon number and the washout of the axion rotation in supersymmetric axiogenesis scenarios.

\section{Rotating axion field}
\label{sec:rotating}

\subsection{Dynamics of rotation}
\label{sec:kick}

In this subsection, we review the rotational dynamics of an axion field proposed in~\cite{Co:2019wyp}.

The PQ symmetry is, at minimum, explicitly broken by the QCD anomaly.
Given that the PQ symmetry is not exact, it is plausible that the symmetry is also explicitly broken by a higher dimensional operator in the superpotential,
\begin{align}
    W = \frac{1}{n}\frac{P^n}{M^{n-3}}.
\end{align}
Here $P$ is a complex field whose angular direction $\theta$ is the axion; its radial direction $S$ is called the saxion. $M$ is a dimensionful parameter. Such a term is expected in theories with an accidental PQ symmetry~\cite{Georgi:1981pu,Holman:1992us,Barr:1992qq,Kamionkowski:1992mf,Dine:1992vx}. Together with $R$ symmetry breaking by the gravitino mass $m_{3/2}$, an explicit PQ-breaking scalar potential is generated,
\begin{align}
    V \sim \frac{m_{3/2}P^n}{M^{n-3}} + {\rm h.c.}
\end{align}

Although the explicit PQ symmetry breaking should be negligible in the present universe with $S=f_a$ to ensure that the solution to the strong CP problem is not spoiled, in the early universe $S$ may take on a large field value so that the explicit PQ breaking is effective. In supersymmetric theories, the possibility of a large field value for $S$ is particularly plausible---in this case the saxion $S$ is a scalar superpartner of the axion, and it may have a flat potential proportional to supersymmetry breaking.
For example, the PQ symmetry may be spontaneously broken by the renormalization group running of the soft mass of $P$ from a positive one in the UV to a negative one in the IR~\cite{Moxhay:1984am}. The potential of $P$ is
\begin{align}
\label{eq:dim_trans}
V(P) = m_S^2 |P|^2 \left( {\rm ln} \frac{2 |P|^2}{f_a^2} -1 \right),
\end{align}
which is nearly quadratic with a saxion mass $\sim m_S$ at $S\gg f_a$.
Another example is a model with two PQ symmetry breaking fields $P$ and $\bar{P}$,
\begin{align}
\label{eq:two_field}
W = X( P \bar{P} - v_{P}^2 ),~~V_{\rm soft} = m_P^2 |P|^2 +  m_{\bar{P}}^2 |\bar{P}|^2,
\end{align}
where $X$ is a chiral field whose $F$-term fixes $P$ and $\bar{P}$ on the moduli space $P\bar{P} = v_{P}^2$.
For $P \gg v_P$ or $\bar{P} \gg v_P$, the saxion potential is again nearly quadratic with $m_S \simeq m_P$ or $m_{\bar{P}}$, respectively. The flat potential allows $S$ to take on a large field value.
For the remainder of the paper, we assume that the saxion potential is nearly quadratic, $m_S^2 S^2/2$, for $S\gg f_a$. When we show constraints on the parameter space, we also assume that $m_S$ is as large as the MSSM scalar masses, $m_S= m_{0}$.

With a large $S$ and the explicit breaking, $P$ receives a kick in the angular direction and begins to rotate. Because of the cosmic expansion, the field value of $S$ decreases and the explicit breaking by the higher dimensional operator eventually becomes negligible.  At this point, $P$ continues to rotate while preserving the angular momentum in field space. Such dynamics of complex scalar fields was considered in the context of Affleck-Dine mechanism~\cite{Affleck:1984fy,Dine:1995kz}. Rotations of the PQ symmetry breaking field initiated by this mechanism are also considered in~\cite{Takahashi:2003db,Kamada:2019uxp}, although the dynamics at later stages discussed below was not considered in these works. It is convenient to normalize the angular momentum, namely the PQ charge, by the entropy density $s$,
\begin{align}
 Y_\theta = \frac{\dot{\theta} S^2}{s},
\end{align}
which is a constant as long as entropy is not produced.

$P$ is thermalized via its interaction with the thermal bath; see Sec.~\ref{sec:therm_saxion}. When this occurs, the radial motion dissipates. The angular motion, on the other hand, remains on account of PQ charge conservation. 
One may wonder whether thermalization causes the angular momentum to be completely converted into particle-antiparticle asymmetry in the thermal bath, but it is free-energetically favored to keep almost all of the charge in the form of the rotation~\cite{Laine:1998rg,Co:2019wyp}.
The resultant motion in field space after thermalization is therefore circular, i.e., has vanishing ellipticity. 

For circular motion, the equation of motion of $S$ requires that $\dot{\theta}^2 = V'(S)/S$. For $S > f_a$, $\dot{\theta} \simeq m_S$. In this phase, the conservation of the charge $\dot{\theta}S^2 \propto R^{-3}$, which can also be derived from the equation of motion of $\dot{\theta}$, implies $S^2 \propto R^{-3}$.  
As $S$ gets close to $f_a$, $\dot{\theta}^2= V'(S)/S$ is no longer constant and begins to decrease. Conservation of charge with constant $S \simeq f_a$ requires $\dot{\theta}$ to decrease in proportion to $R^{-3}$.
To summarize,
\begin{align}
\label{eq:dtheta}
\dot{\theta}  \propto 
\begin{cases}
\ R^0 & S > f_a \\
\ R^{-3} & S\simeq f_a
\end{cases} .
\end{align}

The energy density of the axion rotation scales as
\begin{align}
\label{eq:rho_theta}
\rho_{\theta} \propto 
\begin{cases}
\  R^{-3} & S > f_a \\
\  R^{-6} & S\simeq f_a
\end{cases} .
\end{align}
Because the energy density in the rotation scales as matter for $S>f_a$, it is possible for it to come to dominate over the thermal bath.  In this case,  an epoch of kination domination occurs once $S\simeq f_a$~\cite{Co:2019wyp}.
This indeed occurs if
\begin{align}
\label{eq:Ytheta_AK}
    Y_\theta > 40 \left(\frac{f_a}{10^9\GeV}\right)^{ \scalebox{1.01}{$\frac{1}{2}$} } \left( \frac{100 \TeV}{m_S} \right)^{ \scalebox{1.01}{$\frac{1}{2}$} } \left(\frac{228.75}{g_*}\right)^{ \scalebox{1.01}{$\frac{1}{4}$} },
\end{align}
and the universe becomes radiation-dominated again for temperature below
\begin{align}
\label{eq:TKR}
    T_{\rm KR} = \frac{3 \sqrt{15}}{2 \sqrt{g_*}\pi } \frac{f_a}{Y_{\theta}} \simeq 3 \times 10^6\GeV \left( \frac{f_a}{10^9 \GeV} \right) \left(\frac{40}{Y_\theta}\right) \left( \frac{228.75}{g_*(T_{\rm KR})} \right)^{ \scalebox{1.01}{$\frac{1}{2}$} }.
\end{align}
The matter- and kination-dominated eras enhance the spectrum of primordial gravitational waves, yielding a unique signature~\cite{Co:2021lkc,Gouttenoire:2021wzu}. See~\cite{Giovannini:1998bp,Giovannini:1999bh,Giovannini:1999qj,Riazuelo:2000fc,Sahni:2001qp,Tashiro:2003qp,Boyle:2007zx,Cui:2017ufi,Cui:2018rwi,Auclair:2019wcv} for earlier works on the modification of gravitational wave spectrum by kination domination.

We briefly comment on the axino mass. In what follows, we assume the axino is not the LSP.  This is easily realized in the two-field model in Eq.~(\ref{eq:two_field}), where the axino has a mass as large as the gravitino and can rapidly decay.  For the logarithmic potential shown in Eq.~(\ref{eq:dim_trans}), without any additional ingredients the axino obtains only a loop-suppressed mass and is likely the LSP. Since we assume a thermalized PQ symmetry breaking field, the axino should also be thermalized.  This could lead to overproduction of axinos (as dark matter or as  late-decaying particles). To avoid this, the axino should receive an additional mass, e.g., from a term in the K\"ahler potential $\sim P P^{\prime}$, with $P^{\prime}$ an extra PQ-charged chiral multiplet.%
\footnote{While we do not consider it further, an RPV axiogenesis scenario with an axino LSP could also be viable. For example, the axino LSP could decay before BBN. We find that this is possible in the DFSZ model with a large enough $LH_u$-type RPV and relatively small $f_a$. Another possibility is to have a stable axino LSP with a mass below ${\cal O}(10)$ eV, which would remove cosmological constraints.
To realize such a light axino requires the saxion mass much below the typical scalar mass scale; see the discussion of the dimensional transmutation potential in Ref.~\cite{Co:2021rhi} for details.}

\subsection{From rotation to baryons: charge transfer, axiogenesis, and washout}
\label{sec:axiogenesis}

In this section, we review axiogenesis, the mechanism by which the angular motion of the axion field may be converted to a baryon asymmetry. The angular momentum of the axion field, namely, the non-zero PQ charge, is partially transferred into a quark chiral asymmetry via the strong sphaleron process.  This is further transferred into a lepton chiral asymmetry and a Higgs number asymmetry by the SM Yukawa interactions. Those asymmetries may be transformed into a baryon asymmetry if there exists a baryon number violating process~\cite{Co:2019wyp}; see Fig.~\ref{fig:schematic_RPV}.

Within the SM, the baryon number violation is provided by the electroweak sphaleron process~\cite{Klinkhamer:1984di,Kuzmin:1985mm}. Before the electroweak phase transition, the electroweak sphaleron process is effective, and the baryon asymmetry reaches the thermal equilibrium value given by
\begin{align}
n_B \simeq c_B \dot{\theta} T^2,
\end{align}
where $c_B$ is a constant that is typically $\mathcal{O}(0.1)$. A formula for $c_B$ as a function of axion-SM particle couplings is given in Refs.~\cite{Domcke:2020kcp,Co:2020xlh}.
The baryon asymmetry freezes out upon electroweak symmetry breaking when sphaleron processes become ineffective. For the SM, this occurs at $T\simeq 130$ GeV~\cite{DOnofrio:2014rug}. The final baryon asymmetry is given by~\cite{Co:2019wyp}
\begin{align}
\label{eq:axiogenesis_min}
    Y_B \simeq 8.5 \times10^{-11} 
    \left( \frac{c_B}{0.1} \right) 
    \left( \frac{Y_\theta}{500}  \right)
    \left(\frac{10^8~{\rm GeV}}{f_a}\right)^2.
\end{align}
As we will see in Sec.~\ref{sec:kinetic}, this minimal contribution is smaller than the observed baryon asymmetry after enforcing the requirement that the KMM should not overproduce axion dark matter. This shortcoming may be remedied by the additional baryon number violation provided by RPV as discussed in Sec.~\ref{sec:RPV}.

The quark chiral symmetry is explicitly broken by Yukawa couplings. In combination with the QCD anomaly, which explicitly breaks the PQ symmetry and the chiral symmetry of colored particles down to a linear combination of the two, the symmetry is completely broken.  If this breaking were too strong, the above axion rotation could be washed out and the estimation of the baryon asymmetry would be modified. This could also disrupt the production of dark matter that is discussed in the next subsection. However, all of the chiral symmetries must be broken for washout to occur. Moreover, not all the PQ charge is stored in the form of the chiral asymmetry---the amount susceptible to washout by the chiral symmetry breaking is suppressed by this fraction: $T^2/S^2$.  This leads to an additional suppression of the washout rate. This suppression, when combined with the smallness of the up Yukawa coupling, is sufficient to ensure that the axion rotation is not washed out in the minimal axiogenesis scenario~\cite{McLerran:1990de,Co:2019wyp}.

In the case with supersymmetric particles as considered here, the washout rate may be enhanced or suppressed. The flavor mixing between squarks provides chiral symmetry breaking, but the gluino provides an extra chiral symmetry. If the flavor mixing is large, then it is possible that the chiral symmetry instead is violated at a rate proportional to the Yukawa couplings of the heavier generations.  Alternately, the presence of the extra chiral symmetry (from the gluino) provides a way to protect the PQ charge, even in the presence of the Yukawa couplings. We discuss the interplay of these constraints in Appendix~\ref{sec:flavor} and obtain a lower bound on $f_a$ or an upper bound on scalar mixing. We find that for sufficiently large values of the squark mixing, consistent with present bounds on flavor changing neutral currents (FCNCs), washout may occur for $f_a \lsim 10^9$ GeV; see right panels of Fig.~\ref{fig:washout}. Because FCNCs are suppressed at high scalar masses, the requirement that washout is avoided can be considered the most stringent bound on squark mixing in this case.

In deriving this constraint on the mixing and $f_a$, we fixed the PQ charge so that the axion rotation also explains axion dark matter as described in the next subsection, but the estimation of the washout rates is applicable to a generic PQ charge.  Our analysis on the washout provides a basis to understand axion rotations in supersymmetric models.  Since the flat saxion potential essential for the initiation of the axion rotation is naturally realized in supersymmetric theories, such an analysis is especially welcome.

\subsection{From rotation to dark matter: kinetic misalignment mechanism}
\label{sec:kinetic}

As the saxion settles to the minimum at $f_a$, the axion field's initial kinetic energy $\rho_\theta =\dot\theta^2 f_a^2/2$ may exceed the potential energy barrier with height $2 m_a^2 f_a^2$.  In this case the axion field continues to rotate coherently. Although the rotation slows once the saxion has settled to the minimum according to Eq.~(\ref{eq:dtheta}), it is possible that the rotation is still significant at the time that the axion would begin its oscillation in a conventional misalignment scenario, i.e., when $m_a(T) \simeq H$.  In such cases the axion abundance comes from the kinetic energy of the axion field, which is called the kinetic misalignment mechanism (KMM)~\cite{Co:2019jts},
rather than the potential energy as in the conventional misalignment mechanism~\cite{Preskill:1982cy, Dine:1982ah,Abbott:1982af}.
Given that $\dot\theta (T)$ redshifts according to Eq.~(\ref{eq:dtheta}) and $m_a(T)$ increases when the temperature approaches the QCD confinement scale, the original picture proposed in Ref.~\cite{Co:2019jts} is that the axion rotation halts and the oscillations around its minimum begin when the kinetic energy falls below the potential energy. However, as discussed in Refs.~\cite{Jaeckel:2016qjp, Berges:2019dgr, Fonseca:2019ypl,Morgante:2021bks} for monodromic axion potentials, the anharmonicity of the axion potential leads to the production of axion fluctuations via parametric resonance~\cite{Dolgov:1989us, Traschen:1990sw, Kofman:1994rk, Shtanov:1994ce, Kofman:1997yn}, fragmenting the coherent axion rotation into axion fluctuations.
The effective production rate is estimated as~\cite{Fonseca:2019ypl}
\begin{equation}
    \Gamma_{\rm PR} \simeq \frac{m_a^4(T)}{\dot\theta^3(T)} ,
\end{equation}
and the energy of each produced axion is around $\dot\theta/2$. 
If the KMM is at work, this process becomes important ($\Gamma_{\rm PR} > H$) before the kinetic energy falls below the potential barrier, so it must be taken into account. For the QCD axion, parametric resonance becomes effective around the QCD phase transition because of large $\dot{\theta}$ and the strong suppression of $m_a(T)$ at high temperatures.
The final yield of the axion can be estimated as follows~\cite{Co:2021rhi}
\begin{align}
Y_a \simeq \frac{\rho_\theta}{s \, \dot\theta/2} = \frac{\dot\theta f_a^2}{s} = Y_\theta,
\end{align}
with $s$ the entropy density, and thus the axion yield coincides with the yield of the PQ charge associated with the rotation.
The axion abundance then reads
\begin{align}
\label{eq:KMM}
\frac{\rho_a}{s} = m_a Y_\theta 
\simeq 0.4 \eV 
\left( \frac{Y_\theta}{7} \right)
\left( \frac{10^8~{\rm GeV}}{f_a} \right) .
\end{align}
In this estimate, the number-changing scattering of axions after the parametric resonance production is neglected.
With a kinetic theory~\cite{Zakharov:1985,Zakharov:1992,Micha:2004bv}, one can show that the scattering rate is actually comparable to the Hubble expansion rate just after the production rate becomes comparable to the Hubble expansion rate.
Since the axions are in an over-occupied state, the number-changing scattering can reduce the number density of the axions.  However, the scattering rate drops rapidly as the number density is reduced by the Hubble expansion or number-changing scatterings, so the reduction is expected to be only by an $\mathcal{O}(1)$ factor. Related discussion appears in Refs.~\cite{Micha:2002ey,Micha:2004bv,Co:2017mop}.

From Eqs.~(\ref{eq:axiogenesis_min}) and (\ref{eq:KMM}), one can see that if the observed baryon asymmetry is explained by the minimal axiogenesis, axion dark matter is overproduced unless $f_a \lesssim 10^6$ GeV, which is disfavored by astrophysical constraints~\cite{Ellis:1987pk,Raffelt:1987yt,Turner:1987by,Mayle:1987as,Raffelt:2006cw,Payez:2014xsa,Bar:2019ifz}. To explain the baryon asymmetry from axion rotation, a more efficient channel of the production of a baryon asymmetry is required. As we will show in Sec.~\ref{sec:RPV}, this can be accomplished by RPV. See~\cite{Co:2020jtv,Harigaya:2021txz,Chakraborty:2021fkp} for other proposals.

It is convenient to rewrite the axion abundance in the following way,
\begin{align}
\label{eq:KMM2}
\frac{\rho_a}{s} = m_a Y_\theta = m_a \frac{m_S f_a^2}{\frac{2\pi^2}{45}g_* T_S^3} ,
\end{align}
where $T_S$ is the temperature at which $S$ settles to $f_a$. By requiring Eq.~(\ref{eq:KMM2}) to explain the observed dark matter abundance $\rho_{\rm DM}/s \simeq 0.44 \eV$, one obtains
\begin{equation}
\label{eq:TS_KMM}
T_S \simeq 200 \TeV 
\left( \frac{m_S}{\rm TeV} \right)^{ \scalebox{1.01}{$\frac{1}{3}$} } 
\left( \frac{f_a}{10^8 \GeV} \right)^{ \scalebox{1.01}{$\frac{1}{3}$} }
\left(\frac{228.75}{g_*(T_S)}\right)^{ \scalebox{1.01}{$\frac{1}{3}$} }.
\end{equation}
That is, if axion dark matter is provided by the KMM, this allows us to fix $T_S$, which determines the scaling transition of $\dot\theta$ based on Eq.~(\ref{eq:dtheta}). Eq.~(\ref{eq:TS_KMM}) may be also understood as a lower bound on $T_S$ for axion dark matter not to be overproduced by the KMM.  In what follows, we will see that $T_{S}$ is important both for understanding possible washout effects and the production of the baryon asymmetry.

\subsection{Thermalization of the Peccei-Quinn symmetry breaking field}
\label{sec:therm_saxion}
If unthermalized, the energy associated with radial motion (i.e., the saxion), can have undesirable consequences.  For example, when the saxion ultimately decays, it could produce unacceptable amounts of axion dark radiation that is excluded by the observations of the cosmic microwave background,
or alternately, the entropy produced in its decay might dilute the dark matter or baryon abundance to an unacceptably low level.  In this section, we discuss how thermalization may avoid these effects and how the requirement of thermalization places constraints on the theory.

We first derive a constraint on the maximum $Y_{\theta}$ as a function of the thermalization temperature.  At the time of thermalization, we have $m_{S}^2 S_{\rm th}^{2} \le (\pi^2 g_*/{30})  T_{\rm th}^4$.   Equality holds when the rotation has $\mathcal{O}(1)$ ellipticity and comes to dominate the energy density of the universe prior to thermalization.  On the other hand, at the time of thermalization, we have $Y_{\theta} = m_{S} S_{\rm th}^{2}/(\frac{2\pi^2}{45}g_* T_{\rm th}^3). $ Taken together, these give
\begin{align}
\label{eq:Y_saxMD}
    Y_\theta \le \frac{3 T_{\rm th}}{4 m_S}.
\end{align}
 For a fixed $T_{\rm th}$, Eq.~(\ref{eq:Y_saxMD}) is the maximum achievable yield. 
 One can determine $T_{\rm th}$ by the interaction rate of the saxion with the thermal bath.
 This places a constraint on the parameter space when the yield required for dark matter and/or the baryon asymmetry exceeds this maximum.

The question then becomes what the largest possible $T_{\rm th}$ is for a given $m_S$.  
To answer this question, we first note that $T_{\rm th} > m_{S}$ for the following reasons. 
Constraints on the axion decay constant from supernovae cooling restrict $f_a > 10^8 \GeV$. Taking this bound into account, the dark matter abundance in Eq.~(\ref{eq:KMM}) restricts $Y_\theta > 1$ and thus $T_{\rm th} > m_S$. (Even if one does not assume axion dark matter from the KMM, the values of $Y_\theta$ required by the baryon asymmetry still exceed unity as discussed later in Sec.~\ref{sec:FI}.)

For $T_{\rm th} > m_S$, scattering with the thermal bath is a more efficient thermalization channel than decay.
To provide the necessary interaction, we consider a Yukawa coupling $\mathcal{L} \supset y_\psi P \psi \bar\psi$ where the fermions $\psi, \bar\psi$ are charged under the Standard Model gauge groups and may be the KSVZ quark~\cite{Kim:1979if,Shifman:1979if}. Demanding that the fermions are in the bath at a given temperature requires $y_\psi S < T$.  This leads to the maximal thermalization rate,
\begin{align} 
\label{eq:therm_rate}
    \Gamma_\psi = b y_\psi^2 T \lesssim b \frac{T^3}{S^2} \equiv \Gamma_\psi^{\rm max}, \hspace{1 cm} b \simeq 0.1 \ .
\end{align}
If $\psi$ is not charged, its thermal mass may be much smaller than $m_S$, and the thermalization may proceed via the decay of $S$ into $\psi$, but the rate $\sim 0.1 y_\psi^2 m_S < 0.1 m_S^3/S^2 < \Gamma_\psi^{\rm max}$, where we use $y_\psi S < m_S$.

To determine the maximal possible yield consistent with the thermalization requirement, we use $\Gamma_\psi^{\rm max} = H(T_{\rm th})$, $m_S^2 S_{\rm th}^2 \leq \pi^2 g_* T_{\rm th}^4/30$, and Eq.~(\ref{eq:Y_saxMD}), obtaining
\begin{align}
\label{eq:Y_theta_max}
  Y_\theta^{\rm max} \simeq 10^3 \left( \frac{b}{0.1} \right)^{ \scalebox{1.01}{$\frac{1}{3}$} } 
    \left( \frac{100 \TeV}{m_S} \right)^{ \scalebox{1.01}{$\frac{1}{3}$} } 
    \left( \frac{228.75}{g_*(T_{\rm th})} \right)^{ \scalebox{1.01}{$\frac{1}{2}$} } .
\end{align}

We can now derive the thermalization constraint in the case where the axions produced by the KMM constitute the dark matter. This requires $Y_\theta$ from the KMM in Eq.~(\ref{eq:KMM}) be smaller than the maximum yield in Eq.~(\ref{eq:Y_theta_max}). The bound on the decay constant is 
\begin{align}
\label{eq:fa_max_KMM}
    f_a \lesssim 2 \times 10^{10} \GeV 
    \left( \frac{b}{0.1} \right)^{ \scalebox{1.01}{$\frac{1}{3}$} } 
    \left( \frac{100 \TeV}{m_S} \right)^{ \scalebox{1.01}{$\frac{1}{3}$} } 
    \left( \frac{228.75}{g_*(T_{\rm th})} \right)^{ \scalebox{1.01}{$\frac{1}{2}$} }.
\end{align}
Above this $f_a$ the saxion is thermalized too late to produce sufficient yield for dark matter even with the maximal rate in Eq.~(\ref{eq:therm_rate}). Below this, with an appropriate choice of $y_\psi$, saxion thermalization can occur to allow for the correct $Y_{\theta}$ for the KMM. 

If dark matter has an alternate origin other than the axions produced by the KMM, the thermalization constraint in Eq.~(\ref{eq:Y_theta_max}) can still limit the parameter space.  In this case, information on $Y_\theta$ comes from the requirement that the baryon asymmetry be successfully reproduced.  We will elaborate on this point in Sec.~\ref{sec:FI}.

In the above analysis, we assume that the potential of $S$ is dominated by the vacuum one $\sim m_S^2 S^2$ around and after thermalization. The thermal potential from the coupling $y_\psi$ may dominate in principle, but we find that 
this is not the case, i.e., $y_\psi T_{\rm th} < m_S$, at the time of the thermalization after imposing one of the thermalization constraints $y_\psi S_{\rm th} < T_{\rm th}$.
Note that the thermal potential becomes less and less important at lower temperatures in comparison with the vacuum one, so it is enough to require the consistency at the time of thermalization.%
\footnote{In fact, the requirement is not only for the consistency of the analysis but is a bound~\cite{Co:2021rhi}. If the thermal potential dominates, the potential is flatter than a quadratic one, for which the rotation has instability and Q-balls~\cite{Coleman:1985ki} are formed~\cite{Kusenko:1997zq,Kusenko:1997si,Kasuya:1999wu,Dine:2003ax}. The Q-balls melt once the vacuum potential dominates~\cite{Chiba:2010ff}, but the resultant field configuration is inhomogeneous and needs to be thermalized.}

Since $\psi$ is charged under SM gauge symmetry, its mass should be above ${\cal O}(0.1)$ TeV and ${\cal O}(1)$ TeV for non-colored and colored $\psi$, respectively. This puts a lower bound on $y_\psi$, but we find that the bound is consistent with the upper bound $y_\psi S_{\rm th} < T_{\rm th}$.

If $y_\psi S >T$, the abundance of $\psi$ in the thermal bath is exponentially suppressed.
Still, thermalization may proceed from scattering via the coupling between $S$ and the thermal bath that arises after integrating out $\psi$. Since $\psi$ is charged under Standard Model gauge symmetries, $S$ indeed obtains a one-loop suppressed coupling with gauge fields and is thermalized with a rate $\sim 10^{-5} T^3/S^2$~\cite{Bodeker:2006ij,Laine:2010cq,Mukaida:2012qn}. The corresponding constraint can be obtained by setting $b\sim 10^{-5}$ in the equations above. We have verified that the two-loop thermal logarithmic potential of $S$ generated in this case~\cite{Anisimov:2000wx} is much smaller than the vacuum potential for the saxion masses that we will consider.

\section{RPV axiogenesis}
\label{sec:RPV}

In this paper we focus on dimensionless RPV in superpotential,%
\footnote{We use the ordering of the indices compatible with $SU(5)$ unification, which is different from the standard one in the literature.}
\begin{equation}
\label{eq:RPV}
W = \frac{1}{2}\lambda_{ijk} \bar{e}_i L_j L_k+ \lambda'_{ijk}Q_i L_j \bar{d}_k + \frac{1}{2} \lambda''_{ijk} \bar u_i \bar d_j \bar d_k,
\end{equation}
where $Q_i$, $L_i$, $\bar{u}_i$, $\bar{d}_i$, and $\bar{e}_i$ are doublet quarks, doublet leptons, right-handed up-type quarks, right-handed down-type quarks, and right-handed charged leptons, respectively. 
Rather than investigating the full possible parameter space, we consider the parameter space that is motivated from grand unification. As we will see, in this case the baryon asymmetry is produced by  freeze-in because of the strong upper bound on the magnitude of RPV from proton decay.
RPV is also bounded from below in order for the LSP to decay without disturbing BBN. As a result, some of the parameter space is already disfavored, and the viable parameter region can be further probed by proton decay.

\subsection{Proton decay}
\label{sec:protondecay}

In the context of a grand unified theory such as $SU(5)$, the existence of, e.g., the $\lambda''$ couplings would indicate the presence the $\lambda$ and $\lambda'$ couplings of the similar strength;
\begin{equation}\label{eq:GUTRelation}
    \lambda \simeq \lambda^{\prime} \simeq \lambda^{\prime \prime} \quad{\rm{(GUT \; relation)}} .
\end{equation}
In this case, bounds from proton decay, proportional to the product of $\lambda'' \lambda'$, can be strong.

The strong proton decay constraints limit the size of the RPV within the first generation.  To maximize RPV effects on baryon asymmetry without a large proton decay rate, we assume that the dominant RPV resides in couplings to the $2^{\rm nd}$ and $3^{\rm rd}$ generations.  However, it is unlikely that couplings to the lighter fermions completely vanish, so the constraints can still be significant.  To understand the effects of these residual couplings, we must make assumptions regarding the flavor structure of the RPV.  We assume that the $SU(5)$ ten-plets $(Q,\bar{u},\bar{e})$ are charged under a flavor symmetry and this provides a natural suppression $\lambda_{1jk}^{(','')}\sim \theta_{13}^q \lambda_{3jk}^{(','')}$ and $\lambda_{2jk}^{(','')}\sim \theta_{23}^q \lambda_{2jk}^{(','')}$, where $\theta_{ij}$ represents a typical CKM mixing between the $i^{\rm th}$ and $j^{\rm th}$ generations.  For the five-plets, we consider two cases:
\begin{enumerate}
    \item 
     A hierarchical case, where the five-plets also have flavor structure.  We assume that this flavor structure enforces that only elements with $(j,k)=(2,3)$ and $(3,2)$ are significantly different from zero. This will be the case with the weakest proton decay constraints.  Note that $(j,k)=(3,3)$ identically vanishes in $SU(5)$ unification.
     \item
     An anarchical case, where the five-plets have no flavor structure and $\lambda_{ijk}$ are of  similar  size for all choices of $(j,k)$. The anarchical structure can be motivated by the large mixing angles observed in the neutrino sector~\cite{Haba:2000be}.
\end{enumerate}

\paragraph*{Hierarchical $\bar{d}$ and $L$:}
In this case, the following two couplings,
\begin{align}
    W = \lambda'_{123} Q_1 L_2 \bar{d}_3  + \lambda''_{123}\bar{u}_1 \bar{d}_2\bar{d}_3,
\end{align}
with the exchange of $\tilde{\bar{d}}_3$, generate a dimension-6 operator,
\begin{align}
  & \frac{\lambda'_{123} \lambda_{123}^{''*}}{m_{\tilde{d}_3}^2}Q_1 L_2 (\bar{u}_1\bar{d}_2)^\dag + {\rm h.c.},\nonumber \\
  & \lambda'_{123} \lambda_{123}^{''*}\equiv f_\lambda  \theta_{13}^2|\lambda^{(','')}_{323}|^2,~f_\lambda = {\mathcal O}(1).
\end{align}
Here we have included suppression of $\theta_{13} =4 \times 10^{-3} = |V_{ub}|$ to account for the flavor suppression in the ten-plets and a factor $f_\lambda$ to account for unknown $\mathcal{O}(1)$ factors.
The dimension-6 operator induces $p\rightarrow K^+ \nu$ and $p\rightarrow K^0 \mu^+$.
To compute the proton decay rates, we utilize the lattice calculation of the hadronic matrix elements from~\cite{Aoki:2017puj}, with one-loop renormalization of the Wilson coefficients of the dimension-6 operators from the weak scale down to $2$ GeV as described in~\cite{Buras:1977yy}; we negelect the small effect of the running between the superpartner and weak scales.  The current strongest bound on the proton lifetime on $p\rightarrow K^{+} \bar{\nu}$, $\tau_{K\nu} > 6.6 \times 10^{33}$ years~\cite{Super-Kamiokande:2014otb,Takhistov:2016eqm}, comes from the Super-Kamiokande experiment, which under the present set of assumptions translates to:
\begin{align}
    \lambda^{(','')}_{323} < 2 \times 10^{-9} \left( \frac{m_{\tilde{d}_3}}{10 \TeV} \right) f_\lambda^{-1/2},
\end{align}
which excludes the gray-shaded region in Fig.~\ref{fig:YB_FI}. Here we take $f_\lambda=1$ and $m_{\tilde{d}_3}= m_0$, and use the GUT relation in Eq.~(\ref{eq:GUTRelation}) with $\lambda \equiv \lambda_{323}$.

Future experiments, including DUNE~\cite{DUNE:2018tke} and Hyper-Kamiokande~\cite{Hyper-Kamiokande:2018ofw}, will offer an improvement on the limit on the $p \rightarrow K \nu$ lifetime by roughly a factor of 10.  JUNO will also improve the limit by roughly a factor of 3, perhaps on a shorter time scale~\cite{JUNO:2015zny}.  
Similar bounds on $\lambda$ apply from the $p\rightarrow K^0 \mu^+$ final state.

\paragraph*{Anarchical $\bar{d}$ and $L$:}
With an anarchical structure in the five-plets, the dominant constraint on RPV comes from the following two couplings,
\begin{align}
    W = \lambda'_{21k} Q_2 L_1 \bar{d}_k  + \lambda''_{11k}\bar{u}_1 \bar{d}_1\bar{d}_k. \label{eq:hi_p_decay}
\end{align}
The exchange of $\tilde{\bar{d}}_k$ generates
\begin{align}
  & \frac{\lambda'_{21k} \lambda_{11k}^{''*}}{m_{\tilde{d}_k}^2}Q_2 L_1 (\bar{u}_1\bar{d}_1)^\dag + {\rm h.c.},\nonumber \\
  & \lambda'_{21k} \lambda_{11k}^{''*}\equiv f_\lambda \theta_{13} \theta_{23}  |\lambda^{(','')}_{323}|^2,~f_\lambda = {\mathcal O}(1).
\end{align}
Here we have included suppression of $\theta_{13}= 4\times 10^{-3}$ and $\theta_{23} =0.04 = |V_{cb}|\simeq |V_{ts}| $ to account for the flavor suppression in the ten-plets.
The dimension-6 operator induces $p\rightarrow K^+ \nu$. Following the above-mentioned procedure, we obtain a bound
\begin{align}
    \lambda^{(','')}_{323} < 6 \times 10^{-10} \left( \frac{m_{\tilde{d}_3}}{10 \TeV} \right) f_\lambda^{-1/2},
    \label{eq:an_p_decay}
\end{align}
which excludes the region above the gray line in Fig.~\ref{fig:YB_FI}.
Here we take $f_\lambda=1$ and $m_{\tilde{d}_k}= m_0$.

\subsection{Decay of the LSP}
\label{sec:LSPDecay}
  Because of RPV, the LSP is unstable. Once imposing the above constraints from proton decay, $\lambda$ is required to be small.  This leads to a potentially long lifetime for the LSP.

Moreover, as we will show, the proton decay constraint coupled with the requirement of successful RPV axiogenesis requires a scalar mass above ${\mathcal O}(10)$ TeV. Such large scalar masses are well-motivated in ``without-singlet" scenarios~\cite{Giudice:1998xp,Wells:2003tf,ArkaniHamed:2004fb,Giudice:2004tc,Wells:2004di,Ibe:2006de,Acharya:2007rc,Hall:2011jd,Ibe:2011aa,Arvanitaki:2012ps,ArkaniHamed:2012gw}, where gauginos obtain one-loop suppressed masses by anomaly mediation~\cite{Randall:1998uk,Giudice:1998xp} and one of them is likely to be the LSP. If other one-loop corrections to the gaugino masses are subdominant, the wino is the LSP, which we assume in the following, although the LSP decay rate is of the same order for other gaugino LSPs or the higgsino LSP.
The decay proceeds to three SM fermions via an off-shell sfermion.
The decay rate from $\lambda'_{332}$ and $\lambda'_{323}$, which dominates because of the color factor, is given by~\cite{Dreiner:2008tw} 
\begin{equation}
\label{eq:LSPDecay}
     \Gamma_{\rm LSP} = \frac{g^2  m_{\rm LSP}^5}{1024 \pi^3} \left(
     \lambda^{'2}_{332} \left( \frac{1}{m_{\tilde{Q}_3}^4} + \frac{1}{m_{\tilde{L}_3}^4} + \frac{1}{m_{\tilde{Q}_3}^2 m_{\tilde{L}_3}^2}\right) +
     \lambda^{'2}_{323} \left( \frac{1}{m_{\tilde{Q}_3}^4} + \frac{1}{m_{\tilde{L}_2}^4} + \frac{1}{m_{\tilde{Q}_3}^2 m_{\tilde{L}_2}^2}\right)
     \right)  .
\end{equation}
As a reference point, we take the sfermion masses at $10^{16}$ GeV to be nearly universal, $m_0$, for which $m_{\tilde{Q}_3} \simeq 0.8 m_0$ and $m_{\tilde{L}} \simeq m_0$ at the low energy scale. 

\begin{figure}
\includegraphics[width=0.825\linewidth]{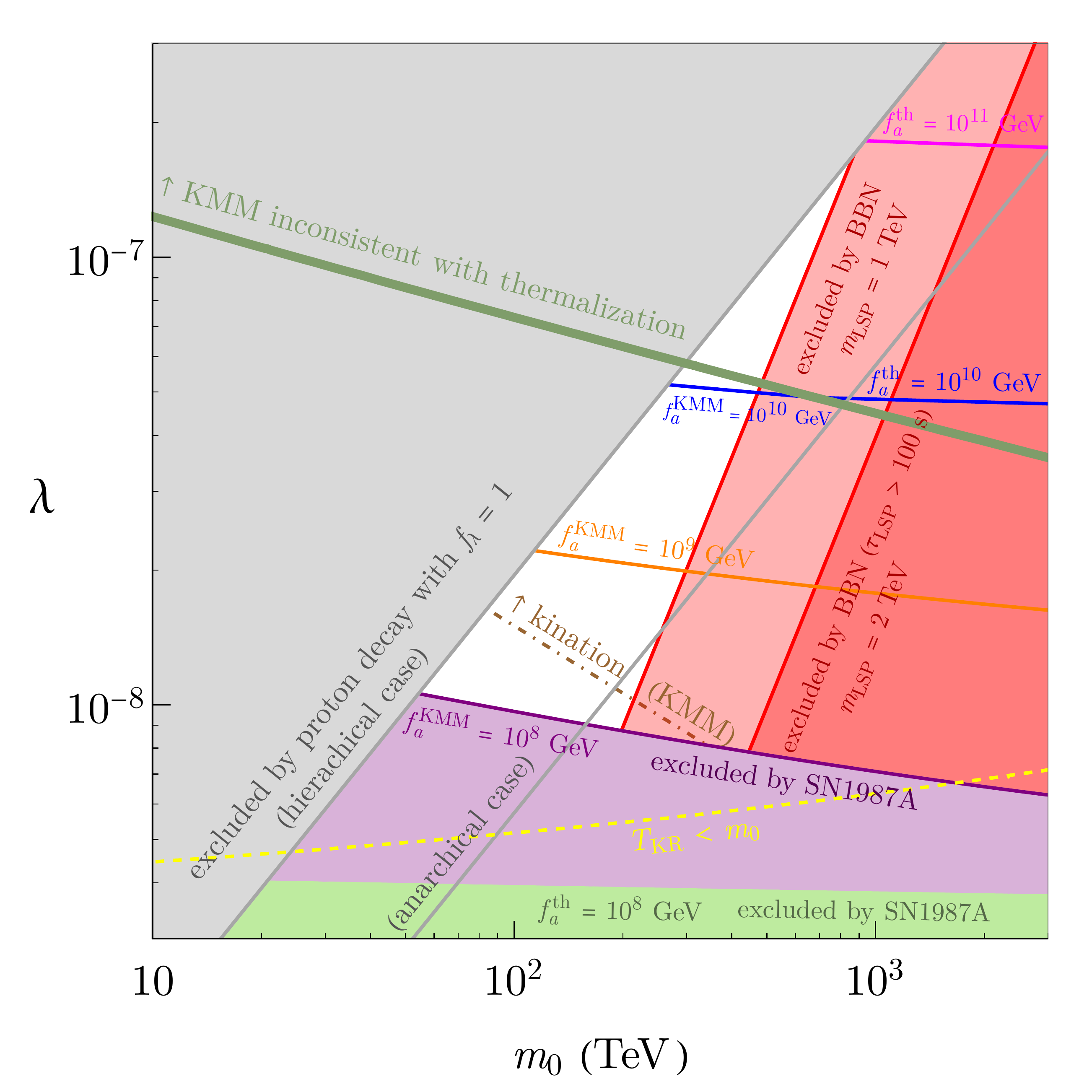}
\caption{
Parameter space of RPV axiogenesis in the scalar mass $m_0$ and the dimensionless RPV coupling $\lambda$.  Light (dark) red shaded region:  late decays of the LSP spoil BBN for $m_{\rm LSP}= 1~(2) \TeV$ (Sec.~\ref{sec:LSPDecay}).
Gray region (line): too rapid proton decay from RPV in the case of hierarchical (anarchical) flavor structure (Sec.~\ref{sec:protondecay}). Solid colored contours: upper bounds on the decay constant $f_a$ required to produce the baryon asymmetry, which equivalently show a lower bound on $\lambda$ for each $f_a$. Those labelled $f_{a}^{\rm KMM}$ predict $f_a$ when dark matter is produced by the KMM or can be interpreted as upper bounds on $f_{a}$ from overproduction. 
 Lines of $f_a^{\rm th}$ give the maximum $f_a$ consistent with thermalization.  
Above the green line, $f_a^{\rm th} < f_{a}^{\rm KMM}$, so it is impossible to achieve thermalization consistent with KMM axion dark matter (Sec.~\ref{sec:therm_saxion}).  Other axion dark matter production mechanisms may be possible. Purple (green) region: SN1987A cooling bound, $f_a \gtrsim 10^8 \GeV$, conflicts with $f_a^{\rm KMM}$ ($f_a^{\rm th}$). 
The KMM contribution may be removed by washout, opening up the purple region. This can be achieved by squark mixing above the yellow line (Sec.~\ref{Sec:no_KMM}). 
Above brown dot-dashed line: kination domination occurs when the KMM explains dark matter (Sec.~\ref{sec:kick}).
}
\label{fig:YB_FI}
\end{figure}

The decay of the LSP must occur without disturbing BBN. The upper bound on the lifetime depends on the abundance of the LSP before it decays. We consider the wino LSP, which annihilates effectively.
For $\mathcal{O}(1-10)$ TeV wino, using the result in~\cite{Kawasaki:2017bqm}, we find that the upper bound on the lifetime is about 100s.  Using Eq.~(\ref{eq:LSPDecay}), we find that in the red-shaded regions of Fig.~\ref{fig:YB_FI}, the lifetime exceeds this limit for $m_{\rm LSP} = 1$ and $2 \TeV$.

Throughout most of the allowed region, the lifetime is quite long, and decays would occur well outside the detector.  In this case, any collider signals are likely to coincide with traditional missing energy searches for supersymmetry.  Owing to the high powers of the supersymmetry breaking parameters that occur in the  lifetime, it can vary rather dramatically.  It is possible that there is a small window of parameters where the scalar mass is relatively light, and the LSP is still kinematically accessible at the LHC, where long-lived decays might conceivably be observable at MATHUSLA~\cite{Lubatti:2019vkf}.  For example, for a wino LSP, at the lower left corner of the white triangle in Fig.~\ref{fig:YB_FI}, $c \tau = 10^5$ km, and at the intersection of the gray and green regions, $c \tau = 10^4$ km. For a higgsino and bino LSP, a similar lifetime applies, but in the case of the bino, its annihilation is less efficient, and the lower bound on the lifetime from BBN is ${\cal O}(0.1)$s.

\subsection{Freeze-in generation of baryon asymmetry}
\label{sec:FI}

In order for the dimensionless RPV to be in thermal equilibrium in the early universe, the $B-L$ violation rate, roughly given by
\begin{align}
\label{eq:Gamma_lambda}
 \Gamma_{B-L} \approx \frac{\tilde{\lambda}^2}{8 \pi} T,    
\end{align} 
must be greater than $H$.  Here, $\tilde{\lambda}$ represents a generic RPV coupling, which could be $\lambda$, $\lambda^{\prime}$, or $\lambda^{\prime \prime}$.
This expression for $B-L$ violation is valid for temperatures above the scalar mass scale $m_0$;  below $m_0$ the rate is exponentially suppressed.
The $B-L$ violation rate given by higher dimensional operators after integrating out sfermions also decreases more quickly than the Hubble expansion rate below $m_{0}$.
The freeze-in regime for $\tilde{\lambda}$, where the RPV interaction is never in thermal equilibrium in the early universe ($\Gamma_{B-L} < H$ at $T \simeq m_0$), is therefore
\begin{align}
    \tilde{\lambda} & \lesssim 2 \times 10^{-6} \left( \frac{m_0}{100 \TeV}\right)^{1/2}. 
\end{align}
With the $SU(5)$ texture we impose, the upper bound on $\lambda$'s from proton decay in Eqs.~(\ref{eq:hi_p_decay}) or (\ref{eq:an_p_decay}) indicates that RPV is never in thermal equilibrium and is in the freeze-in regime unless $m_{0}> 10^9$ or $10^{10}$ GeV.

\begin{figure}
\includegraphics[width=0.49\linewidth]{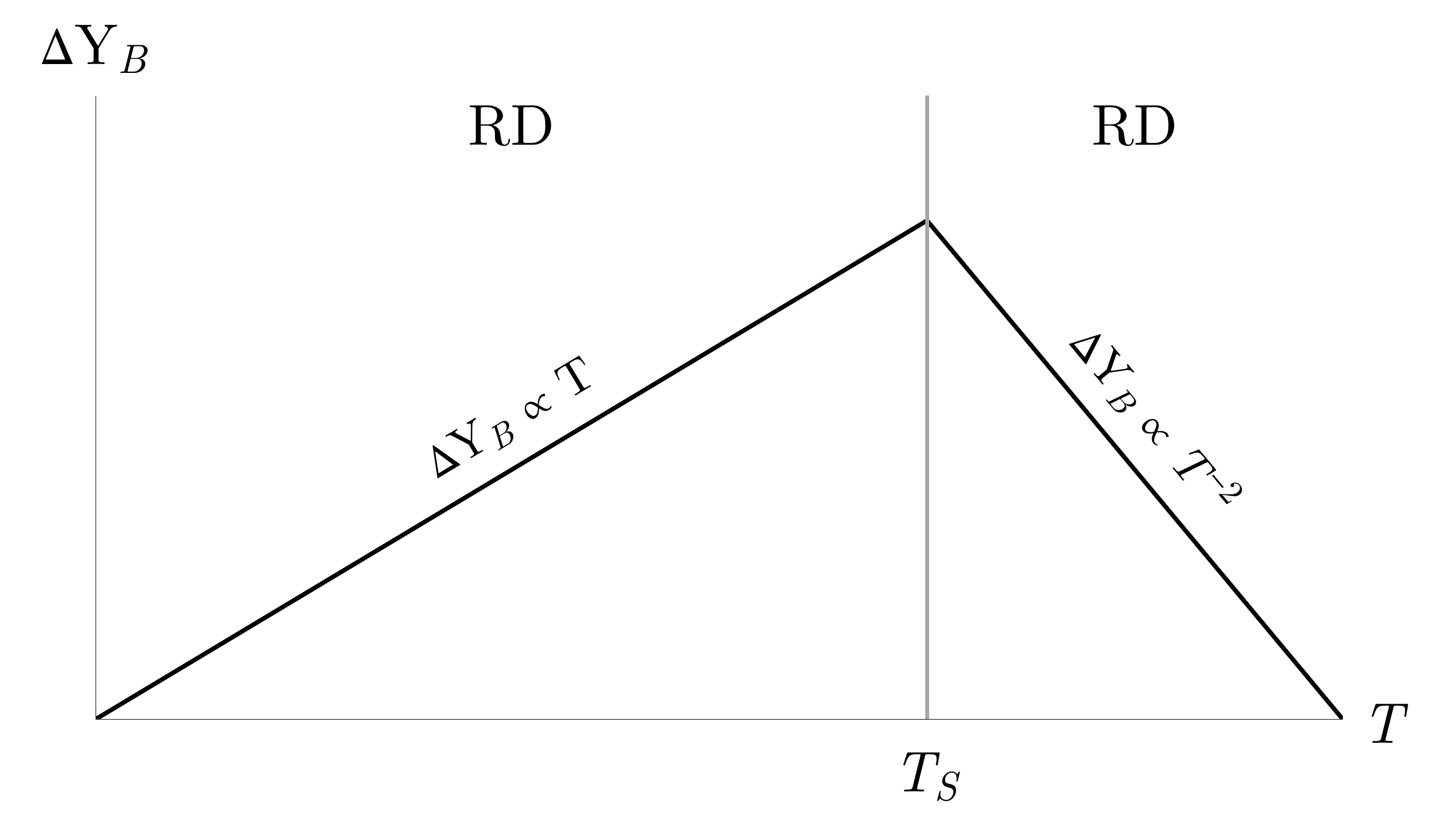}
\includegraphics[width=0.49\linewidth]{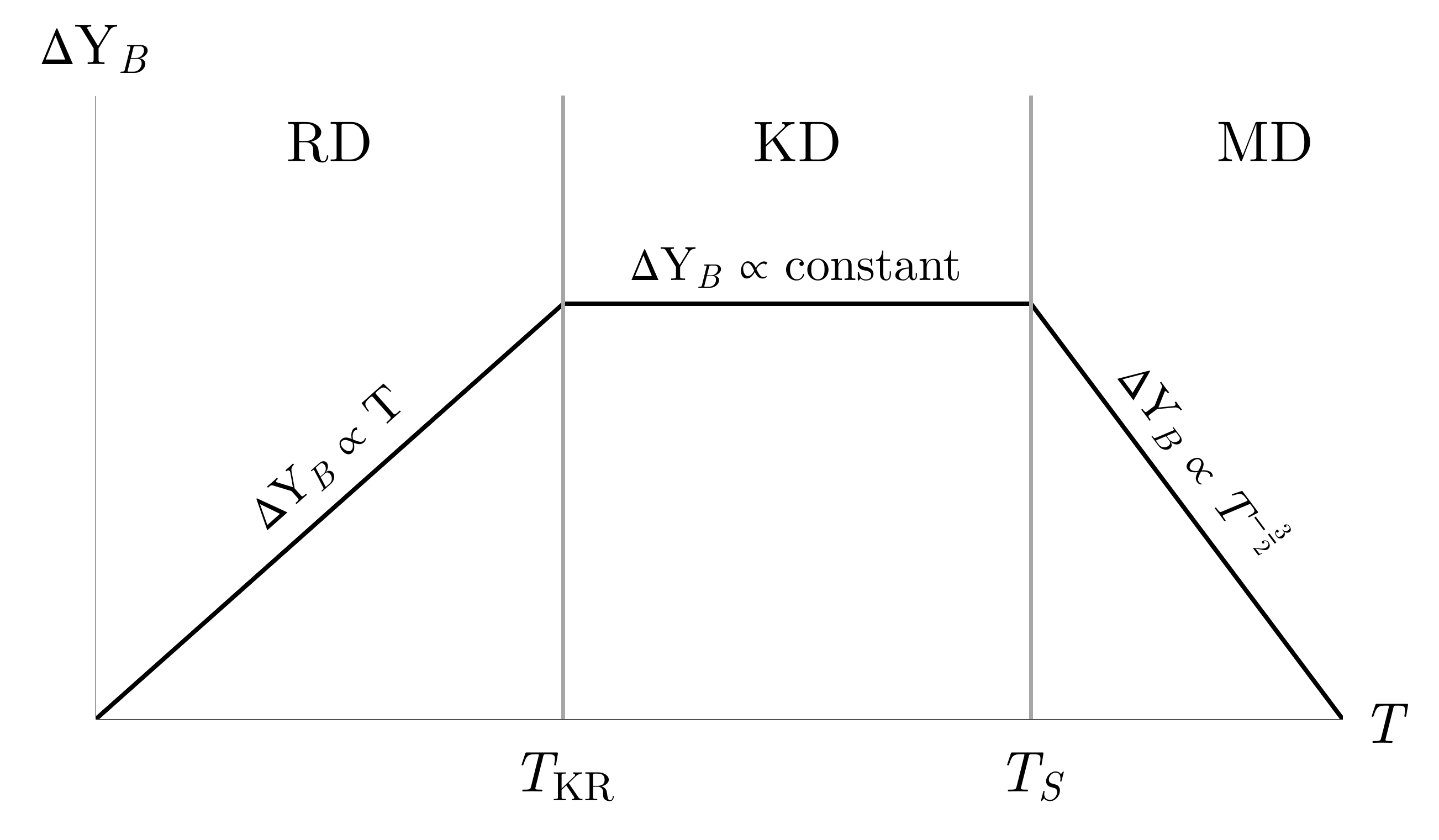}
\caption{The baryon asymmetry produced per Hubble time $\Delta Y_B$ as a function of temperature $T$ in log-log scales during radiation-dominated, kination-dominated, and matter-dominated eras.}
\label{fig:schematic}
\end{figure}

We derive the Boltzmann equation of $B-L$ asymmetry in the freeze-in regime with $T\gg m_0$ in Appendix~\ref{sec:AppRPV}. For the QCD axion that couples to the gluon and weak gauge boson with the same anomaly coefficients, which is the case for the KSVZ model~\cite{Kim:1979if,Shifman:1979if} embedded into grand unified theories, we find
\begin{align}
    \frac{d}{dt}Y_{B-L} = \kappa \lambda^2 \dot{\theta}(T) \frac{45}{2\pi^2 g_*},~~\kappa \simeq 0.007.
\end{align}
Here, to determine the numerical coefficient $\kappa$, we take the dominant contribution from $\lambda''_{332} \equiv \lambda$.
The baryon asymmetry produced per Hubble time is then given by
\begin{align}
\label{eq:YB_FI}
\Delta Y_{B-L} & \simeq \frac{45}{2\pi^2 g_*} \kappa \lambda^2 \frac{\dot{\theta}(T)}{H}~(\sim \frac{\dot{\theta}}{T} \frac{\Gamma_{B-L}}{H}) .
\end{align}
In Fig.~\ref{fig:schematic}, we schematically show the contribution to the baryon asymmetry per Hubble time as a function of temperature above $m_0$.
In the left panel, we show the case where the universe is radiation-dominated (RD) both above and below $T_{S}$, the temperature where the saxion settles to its minimum.
Here we assume $m_0 < T_S$; see the discussion below for the validity of the assumption.
Taking into account the scaling of $\dot{\theta}$ in Eq.~(\ref{eq:dtheta}), the production is peaked at $T_{S}$.  In the right panel, we allow for the possibility that the rotation dominates the energy density of the universe at high temperatures (early matter domination, MD).  In this case, at $T_{S}$, the universe enters a regime where its energy density is dominated by the axion's kinetic energy.  In this epoch of kination domination (KD), $H \propto T^{3}$, and $\Delta Y_{B-L}$ is constant per Hubble time.  This gives an enhancement $\sim \ln ({T_{S}/T_{\rm KR}})$ to the baryon number.  

We approximate the energy density of the axion rotation for $T<T_S$ by $m_S^2 f_a^2 (T/T_S)^6 /2$.
Integrating $\dot{Y}_{B-L}$ over the period with $T\leq T_S$ and multiplying it by $28/79$ to convert $B-L$ to $B$~\cite{Harvey:1990qw}, we obtain the baryon asymmetry produced by RPV axiogenesis,
\begin{align}
\label{eq:YB}
    Y_{B} & \simeq  \frac{28}{79}\frac{45}{2\pi^2 g_*} \kappa \lambda^2 \frac{\sqrt{6} \MPl}{f_a}{\rm arctanh}\left( \sqrt{ \frac{ \frac{1}{2}m_S^2 f_a^2 }{\frac{1}{2}m_S^2 f_a^2 + \frac{\pi^2}{30}g_* T_S^4 }} \right) \nonumber \\
    & \simeq \frac{28}{79}\frac{45}{2\pi^2 g_*} \kappa \lambda^2 \frac{m_S}{H(T_S)} \times 
    \begin{cases}
     1 & : \text{no kination domination} \\
     \ln \left( \frac{T_S}{T_{\rm KR}} \right) & : \text{kination domination}
    \end{cases} .
\end{align}

Using the above calculation of the baryon asymmetry we are in a position to show the allowed region in the new physics parameter space $(m_0,\lambda)$ in Fig.~\ref{fig:YB_FI}.  We assume $m_S = m_0$. A viable range remains after imposing the constraints from proton decay and BBN. As we will now discuss, the QCD axion decay constant $f_a$ is predicted or bounded from the KMM and/or the thermalization of the PQ symmetry breaking field.

The contours show the upper bound on the axion decay constant $f_a$.  Equivalently, they also show lower bounds on $\lambda$ for a given $f_a$.  These upper bounds on $f_a$ are obtained as follows. We first determine $T_S$ upon using Eq.~(\ref{eq:YB}) to explain the observed baryon asymmetry, $Y_B^{\rm obs} = 8.7 \times 10^{-11}$~\cite{Aghanim:2018eyx}, and we can then compute $Y_{\theta} = m_S f_a^2/s(T_S)$ for a given $f_a$. The contours labelled as $f_a^{\rm KMM}$ show an upper bound on $f_a$ that results from requiring $Y_{\theta}$ less than that in Eq.~(\ref{eq:KMM}). This avoids the overproduction of axion dark matter by the KMM. These same contours also show the prediction for $f_a$ should the KMM be the origin of dark matter. Another upper bound of $f_a$, labeled by $f_a^{\rm th}$, is obtained from satisfying the thermalization constraint in Eq.~(\ref{eq:Y_theta_max}) using the $Y_{\theta}$ required by $Y_B$. Here we take $b = 0.1$. These $f_a^{\rm th}$ contours are independent of the origin of the dark matter.  Above the thick green line based on Eq.~(\ref{eq:fa_max_KMM}), the upper bound from successful thermalization $f_a^{\rm th}$ is stronger than $f_a^{\rm KMM}$, so dark matter cannot be explained by kinetic misalignment. Axion dark matter may instead be produced by the conventional misalignment mechanism~\cite{Preskill:1982cy,Dine:1982ah,Abbott:1982af}    
in this region, but for $f_{a} \lsim 10^{11}$ GeV, the misalignment angle after the axion field stops rotating must be tuned to be close to maximal.  

In the purple-shaded region, the upper bound on $f_a$ (determined by overproduction of dark matter from the KMM) is stronger than $10^8$ GeV, which is in contradiction with the lower bound from excessive cooling by axion emission in SN1987A~\cite{Ellis:1987pk,Raffelt:1987yt,Turner:1987by,Mayle:1987as,Raffelt:2006cw,Chang:2018rso,Carenza:2019pxu}. It is possible that a washout, e.g., due to flavor violation in the squark sector, could eliminate the dark matter from the KMM without disturbing the baryon asymmetry. Even in this case, the green region is still excluded by the thermalization constraint and SN1987A. We discuss this in detail in Sec.~\ref{Sec:no_KMM}.

Above the brown dot-dashed line, where Eq.~(\ref{eq:Ytheta_AK}) is satisfied, kination domination by the axion rotation~\cite{Co:2019wyp} occurs, which imprints a unique signature on the spectrum of possible primordial gravitational waves~\cite{Co:2021lkc,Gouttenoire:2021wzu}.  

For the viable range of $f_a = 10^{8\mathchar`-11}$ GeV, the combination of the proton and LSP decay constraints requires $m_0$ between tens of TeV to several hundred TeV for $m_{\rm LSP} = {\cal O}(1)$ TeV. 
It is remarkable that the allowed range of the scalar mass is consistent with ``without-singlet" scenarios~\cite{Giudice:1998xp,Wells:2003tf,ArkaniHamed:2004fb,Giudice:2004tc,Wells:2004di,Ibe:2006de,Acharya:2007rc,Hall:2011jd,Ibe:2011aa,Arvanitaki:2012ps,ArkaniHamed:2012gw}.

In estimating the baryon asymmetry, we assume $\dot{\theta}/T < 1$. The ratio is maximized at $T_S$, so the validity of the assumption requires $m_S < T_S$. Because of the lower bound on $T_S$ in Eq.~(\ref{eq:TS_KMM}), this is always the case for $m_S < 3000$ TeV $(f_a/10^8~{\rm GeV})^{1/2}$.

In the above discussion, we used the $B-L$ violation rate of the form in Eq.~(\ref{eq:Gamma_lambda}). While this form is applicable for RPV couplings of the type $\lambda''\bar{u}\bar{d}\bar{d}$ for any temperatures, for couplings of the type $\lambda'QL \bar{d}$ and $\lambda LL\bar{e}$ the $B-L$ rate can be more subtle.  For these couplings, we may perform a rotation between $(L,H_d)$ to eliminate $\lambda^{(')}$; the result is a superpotential that instead contains $L H_u$.
In fact, at $T> \mu / y_{b,\tau}$, with $\mu$ the higgsino mass parameter, this is a more convenient basis to follow the evolution of the $B-L$ asymmetry~\cite{Davidson:1996cc,Davidson:1997mc},
since the scattering by the bottom and tau Yukawa couplings is more efficient than that by the $\mu$ term,
and a lepton number does not oscillate rapidly if defined in the Yukawa eigenstates.
Correctly using this basis gives a $B-L$ production rate that is suppressed relative to Eq.~(\ref{eq:Gamma_lambda}) at these high temperatures.  However, because we impose 
the $SU(5)$ relation $\lambda\simeq \lambda'\simeq \lambda''$, we may in any case use Eq.~(\ref{eq:Gamma_lambda})---the contribution from $\bar{u}\bar{d}\bar{d}$ dominates.%
\footnote{The $B-L$ production by $\bar{u}\bar{d}\bar{d}$ may be also suppressed at a temperature above the colored Higgs mass, but such a high temperature is irrelevant for RPV axiogenesis.}

\subsection{Washing out KMM axions}
\label{Sec:no_KMM}
As previously mentioned, below the purple line in Fig.~\ref{fig:YB_FI}, axions produced by the KMM will be overabundant once the supernova cooling constraint is applied. To avoid this constraint---and open up parameter space with lower $\lambda$ and $m_0$---requires depleting the KMM contribution. This can be achieved via washout of the PQ charge that sources the KMM, but this must happen in a way that does not reduce the baryon asymmetry.

As long as the axion rotation does not dominate the energy density, such washout could occur between $T_{S}$ and the QCD phase transition. This would leave the baryon asymmetry, dominated by the freeze-in contribution at $T_{S}$, intact. On the other hand, if the axion rotation does dominate at some epoch, it is important for the washout to occur following $T_{\rm KR}$ given in Eq.~(\ref{eq:TKR}), the temperature when the kination-dominated era ends.  This avoids the production of entropy from the washout, which otherwise dilutes the baryon asymmetry.
Note that even below the brown dot-dashed line in Fig.~\ref{fig:YB_FI}, if $Y_\theta$ is above the KMM bound, kination domination can occur.
 As $\lambda$ decreases, $T_{S}$ must decrease to reproduce the desired baryon asymmetry based on Eq.~(\ref{eq:YB}).  This lower $T_{S}$ gives the kinetic energy of the axion a better chance to dominate and induce a kination-dominated era.
Below the dashed yellow line in Fig.~\ref{fig:YB_FI}, a kination-dominated era occurs and ends at
$T_{\rm KR} < m_0$. This means that the washout induced and regulated by squark mixing discussed in Appendix~\ref{sec:flavor} can occur only during the kination-dominated era.  In this case, the washout necessarily produces entropy, which reduces the baryon asymmetry to a value
\begin{align}
    Y_B \simeq \frac{28}{79} \frac{45}{2\pi^2 g_*} \kappa \lambda^2 \frac{\MPl}{f_a} \quad (\text{washout during kination domination}).
\end{align}
Below the yellow line, this is smaller than the observed baryon asymmetry. 
Therefore, for RPV axiogenesis to operate below this yellow line requires an alternate washout mechanism, free from entropy production. Within the MSSM, one possibility is to use chiral symmetry breaking from the $\mu$ term. Another possibility is that a mass of the squark responsible for the washout is smaller than other scalar masses.
As long as these washout conditions are met, it is possible to live in the purple-shaded region, but another mechanism for the generation of dark matter is required. One possibility is discussed in Sec.~\ref{sec:PR}.

However, even if the KMM contribution to dark matter is removed, the thermalization constraints of Sec.~\ref{sec:therm_saxion} will require $f_a$ to not be too large (near the boundary of the purple region in Fig.~\ref{fig:YB_FI}, $f_a^{\rm th} \approx 3 \times 10^8$ GeV). Even if washout occurs in such a way as to avoid disturbing the baryon asymmetry,
the green shaded region remains inaccessible, for in this region $f_a^{\rm th}$ is below the supernova cooling bound.

\subsection{Comments on generic RPV}\label{sec:generic}

We now comment on more generic RPV, leaving a detailed analysis to a future work. For dimensionless RPV, without an $SU(5)$ relation, proton decay constraints are lessened, and the magnitude of RPV may be larger. This opens the possibility that RPV interactions may achieve thermal equilibrium in the early universe, and the $B-L$ asymmetry freezes out once the temperature drops much below the sparticle masses. 

We first discuss how the RPV story changes as a function of the magnitude of $\lambda$ in the case where only a single coupling is present. For fixed sparticle masses and $f_a$, as $\lambda$ increases from the freeze-in regime to this freeze-out regime, the baryon asymmetry changes in the following way. Within the freeze-in regime, the baryon asymmetry increases according to Eqs.~(\ref{eq:Gamma_lambda}) and (\ref{eq:YB_FI}). Once $\lambda$ enters the freeze-out regime, however, the baryon asymmetry drops. This is because the $B-L$ asymmetry produced before RPV is in thermal equilibrium is washed out. For larger $\lambda$, the baryon asymmetry decreases continuously.
There are two values of $\lambda$ that explain the observed amount of the baryon asymmetry, one in the freeze-in regime and another in the freeze-out regime. The freeze-out case predicts a short lifetime of the LSP that may be probed in collider experiments. The lifetime of the LSP is also short enough to be probed in collider experiments even in the freeze-in case if the LSP is a sfermion or is an electroweakino with sfermions not much heavier than the LSP.

Nevertheless, the single coupling picture may be too simple.  There may be several RPV couplings with hierarchical magnitudes. When determining the dominant contribution of various RPV couplings to the baryon asymmetry, it is not as simple as identifying the largest coupling, as we now discuss. This is important: experimental signals are typically controlled by this largest coupling, and a sharp connection with the baryon asymmetry may be lost. 

To analyze the case where several couplings are present, we first need to understand the symmetry structure of the couplings.  The sphaleron transitions and the Yukawa couplings conserve three symmetries, $B/3-L_i (i=e,\mu,\tau)$. The situation is most straightforward to analyze when multiple RPV couplings break only one linear combination of them. This might be the case, for example, if only the $\lambda^{\prime \prime}_{ijk}$ are non-zero.  We first discuss this case, and then move on to the case where the RPV couplings break several different symmetries.

With only one symmetry broken by multiple couplings, the final baryon asymmetry is determined by the largest coupling, whether the individual couplings are in the freeze-in or freeze-out regime. The LSP lifetime is then robustly predicted from the sparticle mass spectrum and $f_a$. We explain this conclusion in two distinct cases. 1) If the largest coupling $\lambda_{\rm max}$ is in the freeze-in region, then automatically other couplings are in the freeze-in region as well. The baryon asymmetry is then determined accordingly by the largest coupling. 2) If the largest coupling $\lambda_{\rm max}$ is in the freeze-out branch, the process associated with $\lambda_{\rm max}$ will keep $Y_B$ at the equilibrium value until $T_{\rm FO}$, a temperature that is determined by $\lambda_{\rm max}$. The interactions from smaller RPV couplings will either freeze out at a higher temperature and are irrelevant in determining the final $T_{\rm FO}$, or make a freeze-in contribution after $T_{\rm FO}$. The freeze-in contribution at $T_{\rm FO}$ is necessarily subdominant because of the freeze-in suppression $\Gamma/H$ in Eq.~(\ref{eq:YB_FI}). Furthermore, since we always have $T_S > m_0$ in our parameter space, $\dot\theta$ is already redshifting as $T^3$ at $T_{\rm FO}$ based on Eq.~(\ref{eq:dtheta}). This implies that $\Delta Y_B$ in Eq.~(\ref{eq:YB_FI}) is UV-dominated and subsequent freeze-in contributions are even more subdominant.  

The picture is more complicated if couplings that break different symmetries ($B/3-L_i$, $B/3-L_j$, $i \neq j$) are present. For concreteness, we consider the case where we have only $\lambda'$ type couplings. Each coupling will break exactly one lepton flavor symmetry.  Let us take two such couplings $\lambda'_1,\ \lambda'_2$, which break $L_1$, and $L_2$, respectively. If they are equal, we recover the situation with a single coupling, except that the baryon asymmetry is doubled, as each coupling can contribute to the generation of $B- \sum_{i} L_{i}$. Now, assume one is larger, with $\lambda'_2 > \lambda'_1$. If the flavors are not mixed by new sources of flavor violation, e.g., by off-diagonal slepton mass matrices, they will evolve independently, and we will obtain separate lepton asymmetries---each contributes to the baryon asymmetry. As mentioned above, the freeze-out abundance is UV-dominated, and thus if both are in the freeze-out regime, $\lambda'_1$ will decouple earlier and give the larger contribution to the baryon asymmetry. If both are instead freeze-in processes, $\lambda'_2$ will set the larger rate and thus the larger $Y_B$. In the case where $L_2$ freezes out, but $L_1$ freezes in, either can be the dominant contribution depending on the exact values of the couplings.
In the presence of sfermion mass mixing, which breaks $B/3-L_i$ into a linear combination, the story becomes yet more complicated, and we will leave details of this for future work. 
However, we note that in the limit of large slepton mixing, individual $B/3-L_i$ are badly broken, and we may simply follow the evolution of total $B-L$, so that the baryon asymmetry is determined by the largest RPV coupling.

Even when multiple symmetries are broken, there are cases where the largest RPV coupling determines the baryon asymmetry through RPV axiogenesis, and the prediction on the LSP lifetime is robust. One possibility is when $L$ is anarchical, for which we expect $B/3-L_i$ breaking of the similar size.   A second, as noted above using the example of $\lambda'$, occurs when the slepton mixing is sufficiently large so that the baryon asymmetry is dominantly produced while the flavors are still well mixed.

Finally, we comment on the possibility that RPV arises from  dimensionful terms in the superpotential, $\mu_i^{\prime} L_i H_u$. Again, such terms are not tied to proton decay and may result in lepton number violation in either the freeze-in or freeze-out regime; the upper bound from the neutrino mass does not exclude the freeze-out regime. In computing the baryon asymmetry, the necessity of the change of the basis discussed in Sec.~\ref{sec:FI} should be taken into account. We leave this possibility, as well as the possibility that both dimensionful and dimensionless couplings are present, for future study.

\subsection{Comments on early parametric resonance}
\label{sec:PR}

We discussed the production of the baryon asymmetry from RPV while taking into account the production of axion dark matter by the KMM, which is a direct consequence of the axion rotation.
However, there are additional axion production mechanisms that may be present in this framework.

Before the completion of the thermalization of the PQ symmetry breaking field, the initial rotation is not circular. At this stage, parametric resonance (PR) production of the fluctuations of the PQ symmetry breaking field may become efficient. This PR is only effective if the ellipticity of the motion of the PQ symmetry breaking field is sufficiently large or if the saxion field value becomes close to $f_a$ by the cosmic expansion; it is then that the rotation experiences the anharmonic part of the potential. Since the field motion becomes circular after thermalization, efficient PR requires sufficiently late thermalization. The precise threshold depends on the details of the saxion potential (larger saxion self interactions can allow more effective PR); see the discussion in the Appendix of Ref.~\cite{Co:2020jtv}.

Bearing in mind the possibility that thermalization may occur sufficiently early so that PR never becomes effective, we now discuss axion dark matter production channels that may be at work if PR is indeed effective.  We comment on how they affect our analysis.

\begin{enumerate}[wide, labelwidth=!, labelindent=0pt]
\itemtitled{Axions from PR}
If axion fluctuations from early PR are not depleted, they also provide axion dark matter whose abundance is comparable to or larger than the KMM abundance when the early stage of the rotation is close to circular or highly elliptic, respectively~\cite{Co:2017mop,Co:2020dya,Co:2020jtv}. The upper bound on $f_a$ from the overproduction of axion dark matter ($f_a^{\rm KMM}$ for the KMM dominated case) becomes then stronger. We find that the resultant axion dark matter, which could conceivably be too warm, is cold enough in the parameter region we consider.  One may wonder if the early PR can explain axion dark matter above the green line in Fig.~\ref{fig:YB_FI}, but we find that the thermalization constraint for this case is still given by Eq.~(\ref{eq:fa_max_KMM}). This is because the relation between $\dot{\theta}S^2$ and $S^2$ in the KMM, which determines the compatibility of the axion dark matter abundance and thermalization, is the same as the relation between $n_{a}$ and $S^2$ in early PR.  

One may wonder that the thermalization of the rotation that we assume simultaneously depletes the PR axions because of the saxion-axion mixing that would be present when the PQ symmetry breaking field is not at the minimum. However, a rotating background is the minimum of the free-energy for a fixed charge, and one of the fluctuation modes around this background may have a Nambu-Goldstone-boson-like nature which could lead to the suppression of its couplings by a derivative. This suppression can potentially prevent the thermalization of such a mode, allowing it to survive to the present day as dark matter. We leave the investigation of the fluctuations around the rotation to future work.

\itemtitled{Axions from cosmic strings}
If the early PR randomizes the PQ symmetry breaking field, cosmic strings are eventually formed~\cite{Tkachev:1995md,Kasuya:1996ns,Kasuya:1997ha,Kasuya:1998td,Tkachev:1998dc,Kasuya:1999hy}. Unlike the usual case without rotations, the axion field value around the cosmic strings rotates. For a domain wall number unity, the cosmic strings (and associated domain walls) decay into axions. With rapid enough rotations, we expect that the string decay is delayed in comparison with the case without rotations and occurs when the late PR (around the QCD phase transition, see Sec.~\ref{sec:kinetic}) becomes effective, so that the axion field stops rotating and the potential energy dominates the dynamics. Because of the delay, the axion abundance from the cosmic strings will be enhanced relative to the case without rotations.  At this time, since the axion mass $m_a(T)$ is larger than the Hubble rate (which determines the axion wavenumber), the energy of axions per quantum is $m_a(T)$, and the amplitude of the axion field is at the most $f_a$, so we expect that the number density of the axions is at the most $m_a(T) f_a^2$. We find that this is smaller than the KMM contribution $\dot{\theta}f_a^2$  by ${\mathcal O}(0.1-1)$ for $f_a = 10^{8\mathchar`-12}$ GeV, and does not affect the constraints shown in Fig.~\ref{fig:YB_FI}. However, the axions produced from the string-domain wall network will have a different spectrum from the KMM axions and may have a different impact on very small-scale dark matter structure. It will be interesting to perform a lattice simulation to investigate the formation of cosmic strings from a rotating PQ symmetry breaking field and the properties of axions emitted from them.

\itemtitled{Axions from long-lived domain walls}
 With a randomized PQ symmetry breaking field, for a domain wall number larger than unity, the resultant cosmic string-domain wall network is stable, so a large enough explicit PQ symmetry breaking must be present to allow the network decay into axions~\cite{Sikivie:1982qv}. After requiring that the explicit breaking not shift the strong CP phase by more than the experimental bound, the network decays after the QCD phase transition, by which time the axion field stops rotating. The estimation of axion abundance assuming no rotations~\cite{Hiramatsu:2010yn,Hiramatsu:2012sc,Kawasaki:2014sqa,Harigaya:2018ooc} is then applicable. This contribution is determined by deep IR dynamics and is independent of the angular momentum of the axion rotation. The observed axion dark matter can be explained without introducing too large a strong CP phase for $f_a \lsim 10^9$ GeV. This can explain axion dark matter in the purple region in Fig.~\ref{fig:YB_FI}, where the KMM contribution must be washed out.  An observable amount of neutron electric dipole moment is predicted unless a CP phase of the theory is fine-tuned. 
\end{enumerate}

To summarize, if the early PR is effective, the first channel (PR axions) can strengthen the upper bound on $f_a$ from the overproduction of axion dark matter, the second channel (rotating string axions) does not affect our analysis, and the third possibility, which exists when the domain wall number is larger than unity, requires $f_a \lsim 10^9$ GeV but can explain axion dark matter even if the KMM contribution is washed out.

\section{Conclusions and discussion}
\label{section:con}
In this paper we proposed a model in which 
the baryon asymmetry of the Universe is generated through supersymmetric RPV interactions and the rotation of the axion. As in the Affleck-Dine mechanism, higher dimensional PQ violating operators deposit energy into the motion of the axion field. The rotational motion is then partially transferred via strong sphaleron processes to give a fermion chiral asymmetry. RPV interactions convert this chiral asymmetry to the baryon asymmetry. The addition of RPV interactions provides sufficient baryon production
without overproducing axion dark matter via kinetic misalignment, a problem for minimal axiogenesis~\cite{Co:2019wyp}.  

We focus on the dimensionless RPV case, and in particular on flavor textures motivated by grand unified theory. With this assumption, all three types of dimensionless RPV interactions are present and approximately equal, yielding stringent proton decay constraints. This predicts sufficiently small dimensionless RPV couplings that the baryon violating interactions never enter thermal equilibrium, freezing in the asymmetry. The precise proton decay bound depends on the RPV flavor structure, and so we examined two extreme cases: an anarchic texture and a hierarchical one.  In the case of hierarchical couplings, proton decay constraints are weaker, but still relevant for constraining the parameter space.  As a corollary, if this model describes nature, imminent signals are possible in up-coming proton decay searches.  In addition, we investigated the effects of RPV induced LSP decay. The LSP must not disturb BBN, which represents an imporant constraint on the model.  LSP decay also provides a possible signal:  it is possible that the decaying LSP may be visible in searches for long-lived particles, perhaps at MATHUSLA~\cite{Lubatti:2019vkf}. The successful thermalization of the PQ symmetry breaking field further constrains the parameter space.

The predictions of this paper are representative  
of axiogenesis scenarios.
Whenever new physics aids the axion rotation in producing the baryon asymmetry,
there is a non-trivial constraint on the parameters of that new physics.    
This is because the mechanism is determined by the parameters of the new physics, the axion decay constant $f_a$, and the angular momentum of the axion field. Enforcing the correct baryon and dark matter density enables solving for the angular momentum and another parameter. All told, we obtain one non-trivial relation among the parameters of the new physics and $f_a$.
In the present work, the new physics parameters are the RPV couplings and the masses of superpartners---quantities that are correlated with proton and LSP decay.  A similar line of reasoning enabled the constraints on the new physics in Refs.~\cite{Co:2019wyp,Co:2020jtv,Harigaya:2021txz,Co:2020xlh}. 

In fact, even without enforcing that the rotation produces all of the dark matter (KMM), simply avoiding the overproduction of dark matter can constrain the parameters as a function of $f_a$.  This specification need not necessarily predict new experimental signals. For example, this requirement only puts a \emph{lower} bound on the masses of new particles in the models presented in~\cite{Co:2019wyp,Harigaya:2021txz}, and possible signals may be pushed to unobservable energies. In RPV axiogenesis with an $SU(5)$ texture, on the other hand, the requirement puts a lower bound on the RPV couplings, which sets a \emph{minimum} proton decay rate. In principle, one could take large scalar masses to suppress proton decay,
but a large scalar mass scale is disfavored for several reasons: the Higgs boson mass, precise gauge coupling unification, and the BBN constraint on the LSP decay, although these constraints can be avoided by ${\rm tan}\beta\lsim 2$, large threshold corrections at the unification scale, and a large LSP mass, respectively. The BBN constraint becomes robust if the LSP is found at the TeV scale at collider experiments.

A $B-L$ asymmetry may be also produced from the dimension-5 Majorana mass term (lepto-axiogenesis)~\cite{Co:2020jtv,Kawamura:2021xpu}. This contribution can give the whole baryon asymmetry for scalar masses above few tens of TeV in the case of degenerate neutrino masses without saxion domination. However, for hierarchical neutrino masses, lepto-axiogenesis is insufficient to generate the baryon asymmetry for scalar masses below few hundreds/thousands TeV without/with saxion domination. The lepto-axiogenesis contribution is also absent if the neutrino masses are of the Dirac type. (See, however, Ref.~\cite{Chakraborty:2021fkp}.)

We commented on the possibility of dimensionful RPV violation and more general dimensionless couplings without the grand unified texture.  Proton decay no longer heavily restricts the model in these cases.  This allows the baryon violating interactions to be large enough to come into equilibrium, and ``freeze-out" RPV axiogenesis to be realized.  In this case,  we find interesting consequences may arise if RPV interactions contribute to asymmetries in different lepton generations, but we leave a detailed consideration to future work.

\section*{Acknowledgments} 
The work was supported by DoE grant DE-SC0011842 at the University of Minnesota (R.C.), the DoE grant DE-SC0007859 (A.P.), and Friends of the Institute for Advanced Study (K.H.).  A.P. would also like to thank the Simons Foundation for support during his sabbatical. The work of R.C.~was performed in part at Aspen Center for Physics, which is supported by National Science Foundation grant PHY-1607611, and was partially supported by a grant from the Simons Foundation.

\appendix

\section{Freeze-in production of $B-L$}
\label{sec:AppRPV}

As discussed in Sec.~\ref{sec:RPV}, once a texture motivated from unified theories is assumed, the $B-L$ asymmetry freezes-in at a temperature $T_S \gg m_S$ ($\sim m_0$.) The production rate can be then estimated working in the supersymmetric limit.

We compute the production rate of $B-L$ asymmetry from
\begin{align}
\label{eq:appRPV}
W =  \lambda \epsilon_{abc} \bar{u}_3^a \bar{d}_3^b \bar{d}_2^c,
\end{align}
which dominates over other dimensionless RPV couplings because of color factors.
We use the Boltzmann approximation for the thermal distribution function, i.e., do not distinguish between fermions and bosons, which we expect to be accurate to ${\mathcal O}$(10\%).  Within this approximation, we may exploit supersymmetry.

\begin{figure}
\includegraphics[width=0.32\linewidth]{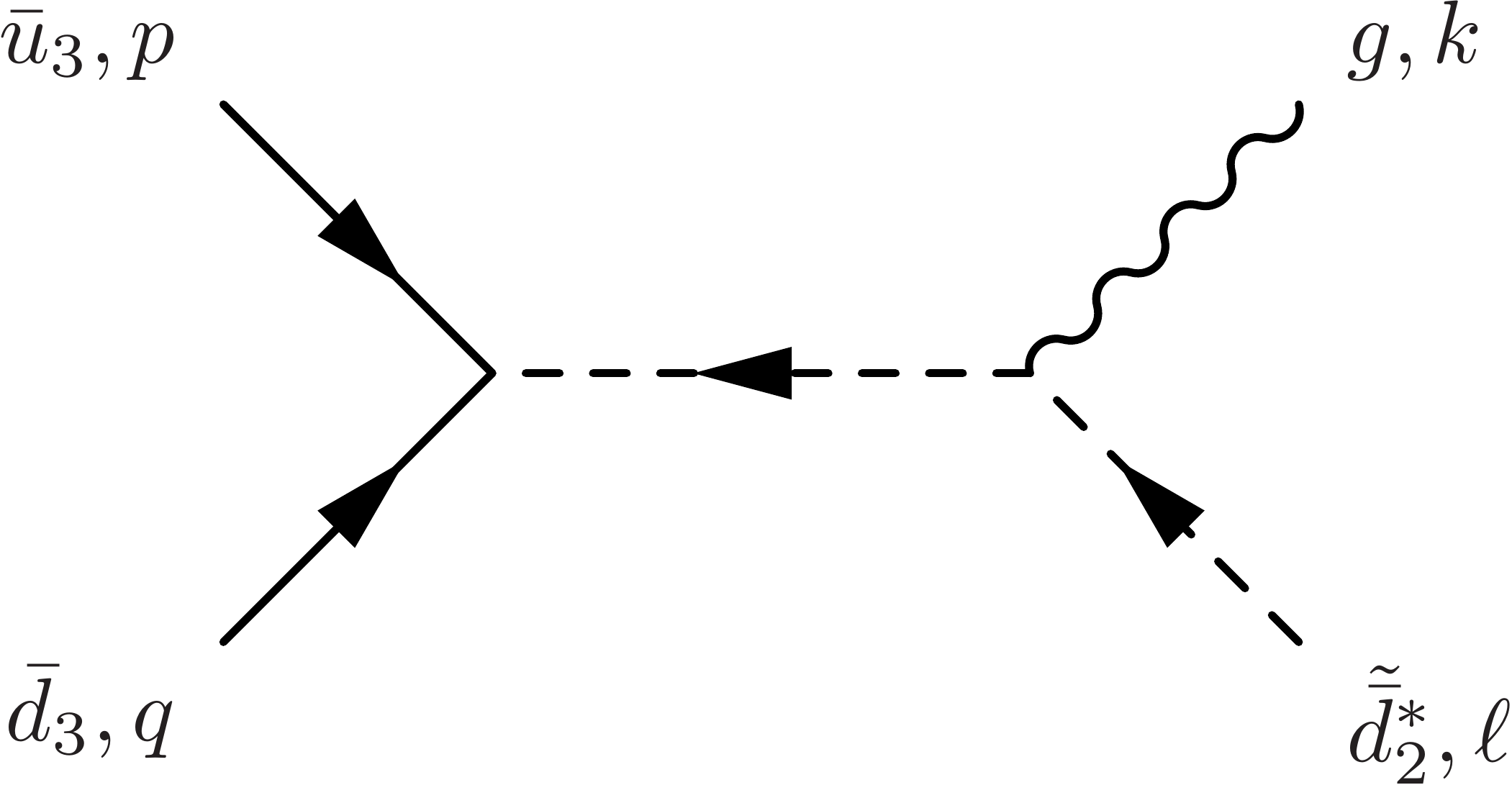}
\includegraphics[width=0.32\linewidth]{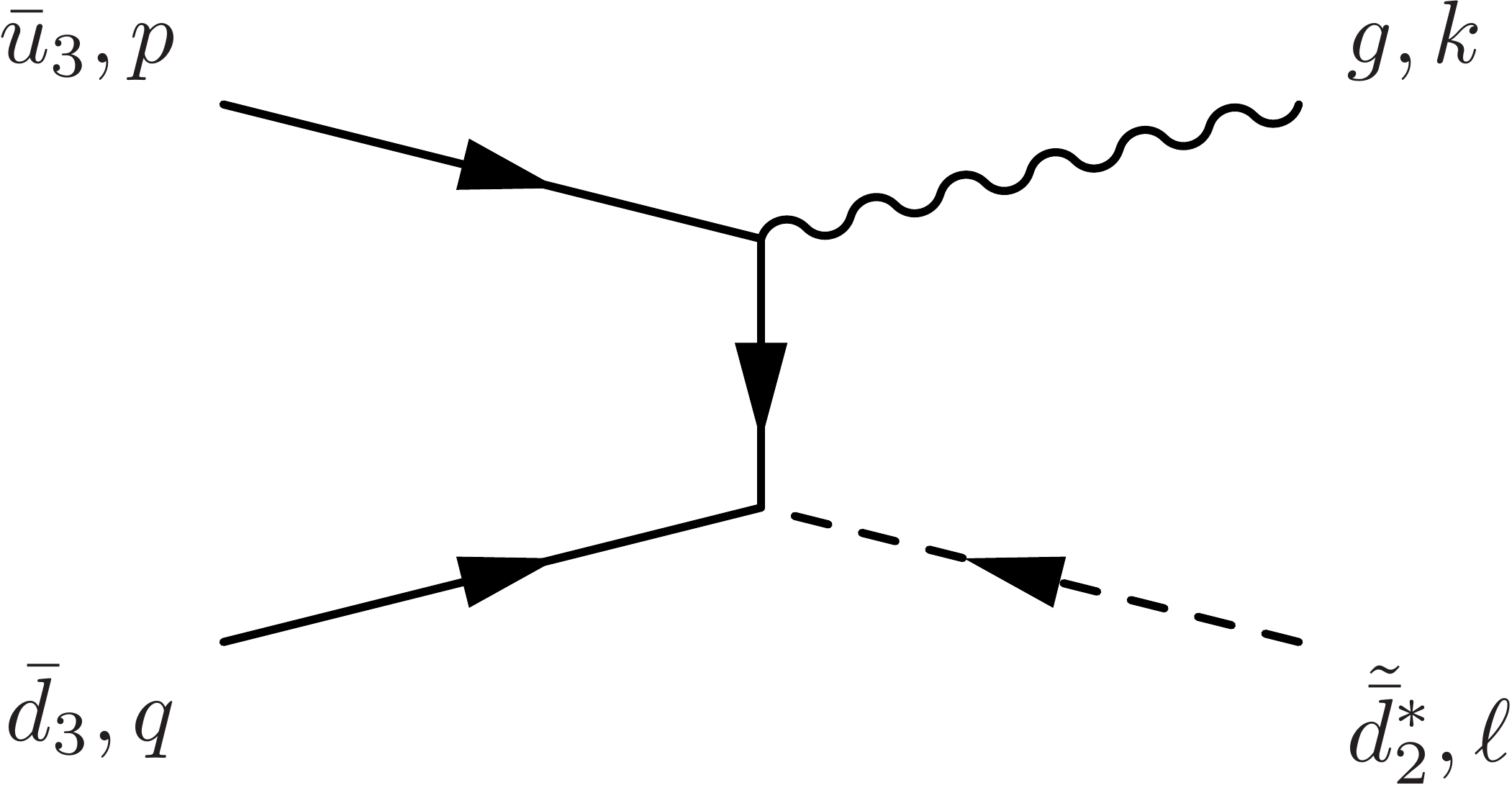}
\includegraphics[width=0.32\linewidth]{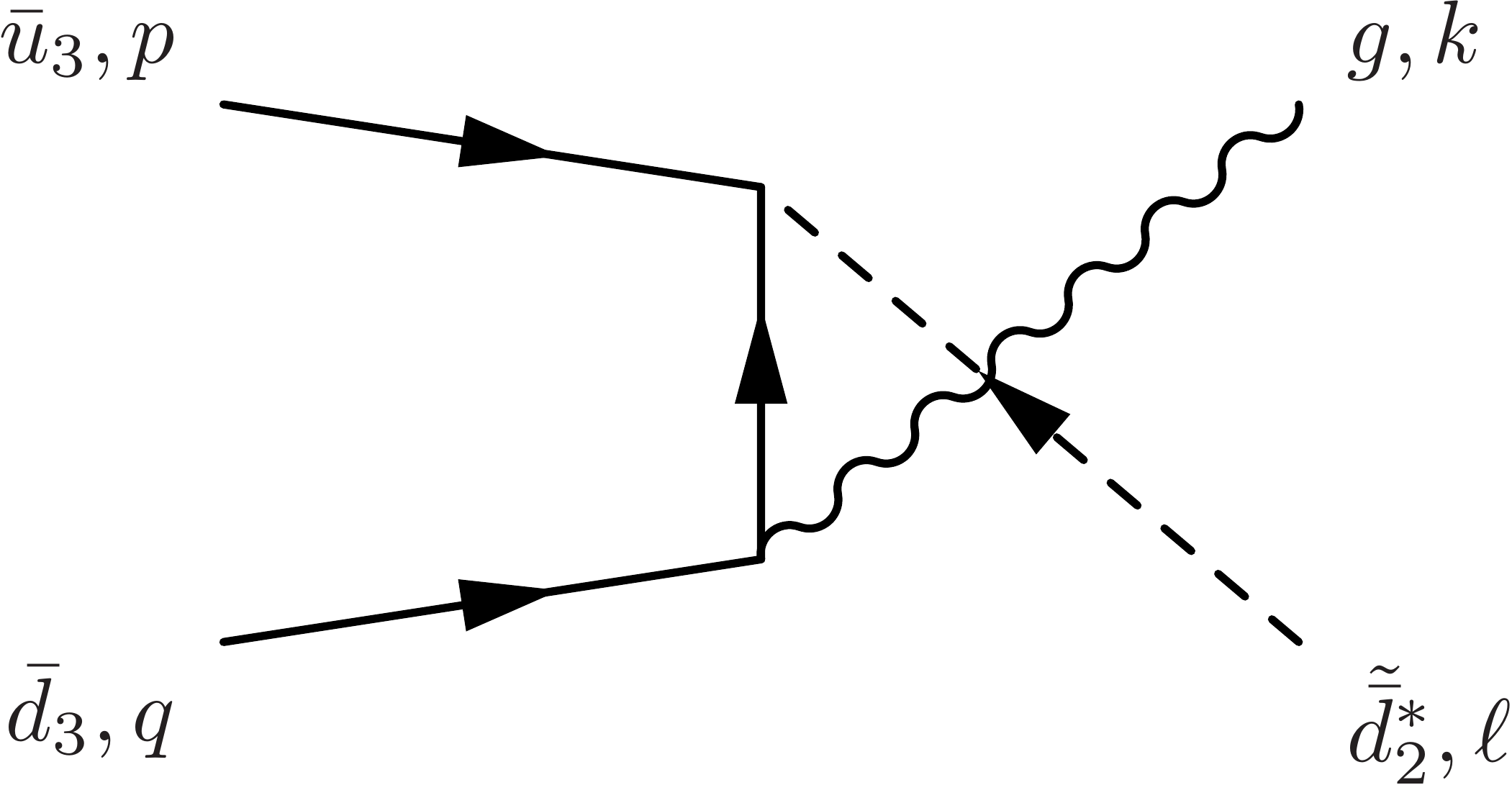}
\includegraphics[width=0.32\linewidth]{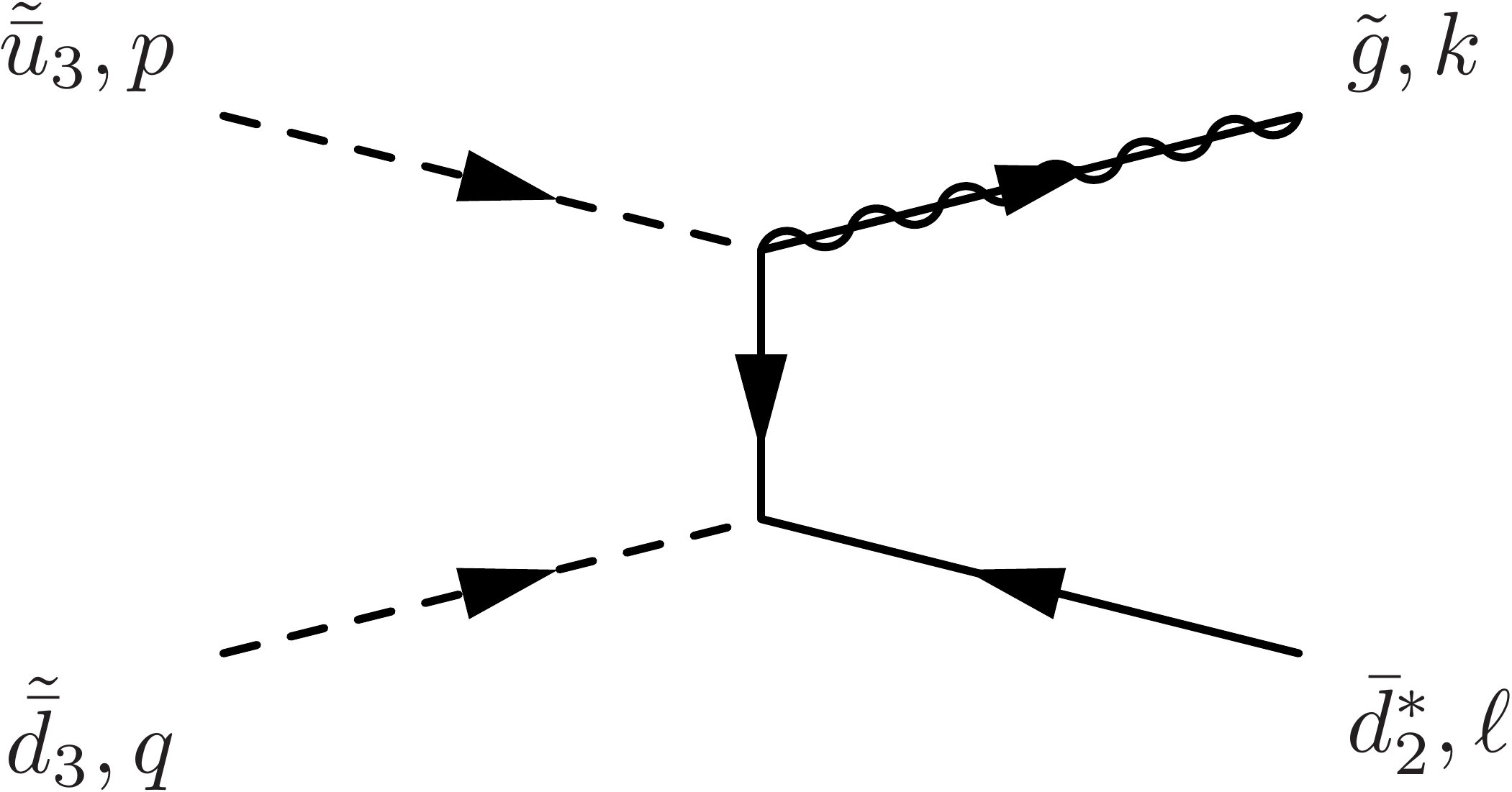}
\includegraphics[width=0.32\linewidth]{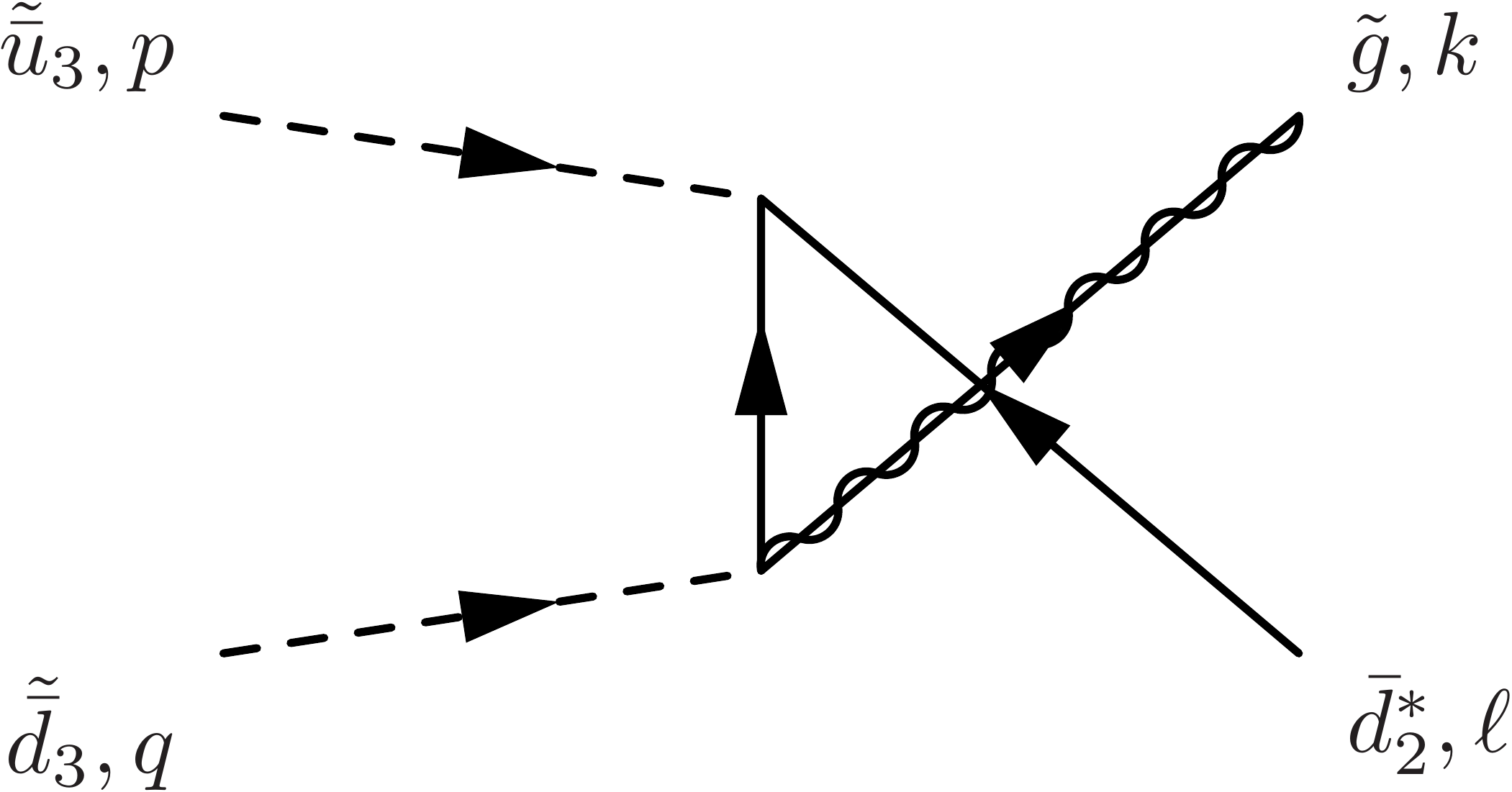}
\caption{$B-L$ violating scattering from a dimensionless RPV coupling.}
\label{fig:diagram}
\end{figure}

With the RPV of Eq.~(\ref{eq:appRPV}), $B-L$ is violated  by the processes shown in Fig.~\ref{fig:diagram}, along with those related by crossing and those with different combination of quarks and squarks.
In the massless limit, the squared matrix elements of the scattering processes are
\begin{align}
\label{eq:Msq}
    \left|{\mathcal M}(\bar{u}_3 \bar{d}_3 \rightarrow \tilde{\bar{d}}^*_2 g)\right|^2_{\rm ave}= \left|{\mathcal M}(\tilde{\bar{u}}_3 \tilde{\bar{d}}_3 \rightarrow \bar{d}^*_2 \tilde{g})\right|^2_{\rm ave} =
    \frac{\lambda^2 g_3^2}{27} \left( 1 + \frac{t}{u} + \frac{u}{t} \right),\nonumber\\
    s \equiv (p+q)^2,~t\equiv (p-k)^2,~u\equiv(p-\ell)^2,
\end{align}
where we take an average over the colors of the external particles and the helicity of the gluon or gluino. The equality of the squared amplitudes of the two processes is guaranteed by a combination of supersymmetry and crossing symmetry.
The amplitudes for the processes involving three quarks and a gluino or three squarks and a gluon may be shown to identically vanish using supersymmetry. Diagrammatically, the former has three tree-level diagrams, but they cancel in supersymmetric limit,%
\footnote{The cancellation should be incomplete for scattering on the thermal background that breaks supersymmetry. The residual scattering rate is suppressed by extra coupling constants and numerical factors smaller than unity, and we neglect it.}
while the latter does not have a tree-level diagram.

Taking the approximation of Boltzmann statistics, the contribution of each process to the $B-L$ production rate is
\begin{align}
    \dot{n}_{B-L} \supset  \int d\Pi_p d\Pi_q d\Pi_k d\Pi_\ell   e^{- \frac{E_p + E_q}{T} } \left(e^{\frac{\mu_p}{T} + \frac{\mu_q}{T}}- e^{\frac{\mu_k}{T} +\frac{\mu_\ell}{T}} \right)     (2\pi)^4 \delta^4(p+q-k-\ell) 
    |\mathcal{M}|^2.
\end{align}
Here $\mu_p$ is the chemical potential for the particle with momentum $p$, and $d\Pi_p=\frac{d^3p}{(2\pi)^3 2E_p}$ is the Lorentz invariant phase space factor.
We assume $\mu \ll T$, which is valid when $\dot{\theta} \ll T$.
Then summing over all the processes gives  
\begin{align}
    \dot{n}_{B-L} =& 3^3 \times 16  \times 3 \times 2 \times \frac{\mu_{\bar{u}_3} + \mu_{\bar{d}_3} + \mu_{\bar{d}_2}}{T} \int d\Pi_p d\Pi_q  d\Pi_k d\Pi_\ell~e^{- \frac{E_p + E_q}{T}}    \\
    &(2\pi)^4 \delta^4(p+q-k-\ell) \frac{\lambda^2 g_3^2}{27}2\left(
    \left(1 + \frac{t} {u} + \frac{u}{t}\right)
    -\left(1 + \frac{s} {u} + \frac{u}{s}\right)
    -\left(1 + \frac{t} {s} + \frac{s}{t}\right)
    \right). \nonumber 
\end{align}
Here the factor $(3^3 \times 16)$ is from the sum over colors and helicities;
$3$ is from the sum over the processes with different combinations of quarks and squarks, and
$2$ is from the sum over two processes in Fig.~\ref{fig:diagram}.  The integrand can be found from Eq.~(\ref{eq:Msq}) after summing over different channels using crossing symmetry.
The integration over $k$ and $\ell$ may be done in the center-of-mass frame. The integration over the polar angle $\theta$ is IR-divergent at $\cos{\theta} = \pm 1$ because of the $t$- and $u$- channel poles, which is cut off by the thermal mass squared of quarks $\sim 0.1T^2$. Since the dependence on the cut-off is only logarithmic, to obtain a numerical value, we simply take the cut-off on the integration of $\theta$ to be $\cos{\theta_*} = \pm (1-0.1) = \pm0.9$.   
We then obtain
\begin{align}
\label{eq:BLrate1}
    \dot{n}_{B-L} \simeq 0.04 \lambda^2 T^3 \left(\mu_{\bar{u}_3} + \mu_{\bar{d}_3} + \mu_{\bar{d}_2} \right) =  0.04 \lambda^2 T \frac{\pi^2}{12}\left(n_{\bar{u}_3} + n_{\bar{d}_3} + n_{\bar{d}_2} \right).
\end{align}
Here the particle-antiparticle asymmetry is defined as the sum over scalars and fermions.

The asymmetries $n_{\bar{u}_3,\bar{d}_3,\bar{d}_2}$ are produced from the PQ charge by the strong and/or electroweak sphaleron processes and the Yukawa interactions. Since these processes are efficient, the asymmetries reach thermal equilibrium. The equilibrium values may be obtained from the Boltzmann equation by taking the time derivative of particle-antiparticle asymmetries to be zero while imposing appropriate conservation laws. The resultant equilibrium values are complex functions of coupling constants. Well-approximated, much simpler equilibrium values can be more easily obtained in the following way. As long as one chiral symmetry of a colored particle is unbroken, the system, including the axion rotation, may be in thermal equilibrium with non-zero asymmetry. We may then apply the standard requirement of chemical equilibrium for each interaction~\cite{Harvey:1990qw}, and the equilibrium value of the various asymmetries (and $\dot{\theta}$) are found to be independent of coupling constants. While we wish to find the equilibrium values for the actual case where the chiral symmetry is completely broken, the leading term is reproduced by taking the value of the ``least-broken" chiral symmetry to zero, for which the above procedure is sufficient. In the MSSM, the least-chiral symmetry breaking parameter is the up-quark Yukawa coupling.
We then find
\begin{align}
\label{eq:BLrate2}
    n_{\bar{u}_3} = \frac{124+ 84 c_W}{79\pi^2}\dot{\theta}T^2,~~n_{\bar{d}_3} = n_{\bar{d}_2} = \frac{-92+ 60 c_W}{79\pi^2}\dot{\theta}T^2.
\end{align}

Using Eqs.~(\ref{eq:BLrate1}) and (\ref{eq:BLrate2}), we find
\begin{align}
    \dot{n}_{B-L} \simeq -0.003 \left(1- \frac{17}{5} c_W \right) \lambda^2 \dot{\theta} T^2.
\end{align}
In this estimation, we assume that the electron Yukawa interaction is in thermal equilibrium. This is not true at high temperatures and the equilibrium values of the asymmetry change accordingly, but we find that the $B-L$ production rate changes only by ${\cal O}(10)$\%.

\section{Flavor and washout}
\label{sec:flavor}

In this Appendix, we estimate the washout rate of the axion rotation in the MSSM.  This requires taking into account both scalar mixing---which can provide additional breaking of flavor symmetries---and the presence of the additional chiral symmetry from the gluino. 

In the presence of an effective sphaleron process, the axion rotation is transferred into a chiral asymmetry of colored fermions. The transport equations are given by  
\begin{align}
    \dot{n}_{\theta} & =  - \frac{\Gamma_{\rm ss}}{T^3} \left( \dot{\theta} T^2 - 6 \sum_i \frac{2 C_i}{d_i} n_i \right), \nonumber \\
    \dot{n}_{i} & =  2 C_i \frac{N_i}{d_i} \frac{\Gamma_{\rm ss}}{T^3} \left( \dot{\theta} T^2 - 6 \sum_j \frac{2 C_j}{d_j} n_j \right) + \cdots.
\end{align}
Here, $i$ denotes a colored fermion and $n_i$ is the asymmetry of the fermion, summed over gauge indices. $C_i$ is a quadratic Casimir invariant of the corresponding $SU(3)_c$ representation ($1/2$ for fundamental, $3$ for adjoint, ...), $d_i$ is the dimension of the representation ($3$ for fundamental, $8$ for adjoint, ...), and $N_i$ counts the degrees of freedom of the fermion (e.g., $3\times 2 =6$ for a doublet quark.) $\Gamma_{\rm ss}$ is the sphaleron transition rate per unit volume and time. We use $\Gamma_{\rm ss} \simeq 100 \alpha_s^5 T^4$ estimated in~\cite{Moore:2010jd}. The coefficients are determined following Sec.~6 of Ref.~\cite{Cline:1995dg} and the detailed balance relation between the rotation and chiral asymmetry~\cite{Co:2019wyp}. The ellipsis denotes other terms.  If QCD sphaleron processes are effective and all chiral symmetries of colored fermions are simultaneously broken by processes  
other than the QCD sphaleron, then washout of the rotation may occur. If the transfer by the QCD sphaleron process is the bottleneck, the washout rate is given by~\cite{McLerran:1990de,Co:2019wyp}
\begin{equation}
\label{eq:washoutSS}
    \gamma_{\rm ss} \equiv  \frac{\dot{n}_\theta}{n_\theta} =  \frac{\Gamma_{\rm ss}}{T^3} \frac{T^2}{S^2} \simeq 100 \alpha_{s}^5 \frac{T^3}{S^2}.
\end{equation}
 Typically, another process is the bottleneck and the washout rate is suppressed.

In the Standard Model, there is a $U(1)$ symmetry associated with right-handed up-quark number only broken by the up Yukawa coupling.  This rate is smaller than $\gamma_{\rm ss}$, so the test of whether washout occurs is whether the rate given by~\cite{McLerran:1990de,Co:2019wyp}  
\begin{equation}
\label{eq:washoutYukawa}
    \gamma_u \sim \alpha_{s} y_{u}^2 \frac{T^3}{S^2}
\end{equation}
exceeds the Hubble scale.
The rate is a constant at $S > f_a$, but decreases in proportion to $T^3$ once $S$ has settled to its minimum, so $\gamma_u/H$ is maximized at $T= T_S$. 
For axion dark matter from the KMM, using Eq.~(\ref{eq:TS_KMM}), we find that the washout does not occur when $m_S \lesssim 10^{15} \GeV (f_a/10^8 \GeV)^5 (10^{-5}/y_u)^6$. Here we assume radiation domination at $T_S$. Thanks to the small up Yukawa coupling, washout is easily avoided.

In the supersymmetric case considered here, there are two new wrinkles.  These can be important for determining whether washout occurs for temperatures near or above the superpartner mass scale.   First, there is a new chiral symmetry associated with the gluino mass.  The rate for violation of this symmetry is proportional to the gluino mass,%
\footnote{Here the rate is larger than the Dirac mass shown in~\cite{Kamada:2016eeb} by a factor of 2 for the following reason.
The chiral charge of each fermion changes by two for scattering by a Majorana mass, but only one for a Dirac mass.  The rate is enhanced by a factor of $2\times2 =4$, where one $2$ simply comes from the change in the chiral charge per scattering, and the other $2$ comes from a twice as large bias from the chemical potential.  On the other hand, a Majorana mass provides only one chirality-changing process $\psi \rightarrow \psi^\dag$ while a Dirac mass provides two processes $\psi \rightarrow \bar{\psi}^\dag$ and $\bar{\psi} \rightarrow \psi^\dag$, which provides a relative suppression of  the Majorana rate by a factor of $1/2$.}
\begin{align}
    \dot{n}_{\tilde{g}} = - \frac{24}{\pi^2} \frac{m_{\tilde{g}}^2}{T^2} \Gamma_{\tilde{g}} n_{\tilde{g}} + \cdots,~~ \Gamma_{\tilde{g}} \simeq 4 \alpha_s T.
\end{align}
Here we used the results derived in~\cite{Lee:2004we,Kamada:2016eeb} with the thermal width $\Gamma_{\tilde{g}}$ computed in~\cite{Braaten:1992gd,Kobes:1992ys}.
The washout rate when the gluino chiral symmetry breaking is the bottleneck is
\begin{equation}
\label{eq:washoutGluino}
    \gamma_{\tilde{g}} \simeq
    0.4 \alpha_s \frac{m_{\tilde{g}}^2}{T} \frac{T^2}{S^2}.
\end{equation}
Here we assume that $m_{\tilde{g}} < T$, and while the above estimate breaks down as $T$ approaches $m_{\tilde{g}}$,  the strong sphaleron rate will be smaller in this limit, $\gamma_{\rm ss} \ll \gamma_{\tilde{g}}$, and the gluino chiral symmetry will not provide the bottleneck.  Incidentally, there is another chiral symmetry in the limit where the higgsino mass $\mu \rightarrow 0$.  The relevant washout rate is $\sim \alpha_2 \mu^2 T/S^2$. However, since we assume $\mu \gg m_{\tilde{g}}$, the process involving the gluino mass is more important for determining whether washout occurs.

The second complication is that supersymmetry breaking famously can introduce new sources of flavor violation, and hence it can lead to breaking of the chiral symmetries no longer proportional to the small up Yukawa coupling. Multiple new sources of flavor violation may be necessary to dramatically enhance the washout rate. Because all chiral symmetries must be broken for washout to occur, even if, e.g., the right-handed up quark number is violated by the mixing of right-handed up squarks, the washout rate is still suppressed by the approximate first-generation left-handed quark number and right-handed down-quark number conservation, and hence is proportional to $y_d^2$.  

Let us consider the mixing of the first and third generation right-handed down squarks given by an off-diagonal mass squared $\tilde{m}^2_{13} \equiv \delta_{13} m_0^2$.
We denote the diagonal elements of the mass squared as $\tilde{m}_{11}^2$ and $\tilde{m}_{33}^2$.
One may compute the flavor violation rate in two different bases: 1) the basis where the down quark Yukawa couplings are diagonalized (\emph{Yukawa basis}), and 2) the basis where the scalar masses are diagonalized (\emph{scalar mass basis}).  To determine the rate of flavor symmetry breaking, one should take the basis with a smaller rate, since the stronger interaction defines the good flavor basis to begin with~\cite{Davidson:1996cc,Davidson:1997mc}. The weaker interaction causes flavor violation in that basis. In the Yukawa basis, flavor changing is induced by the off-diagonal scalar mass squared with a rate
\begin{align}
\label{eq:Yukawabasis}
    \Gamma_{13,{\rm y}} \sim \alpha_s \delta_{13}^2 \left( \frac{m_0^4 T^3}{(T^2 + m_0^2)^3}  f_0(T, m_0) + \frac{m_0^4 T^5}{(T^2 + m_0^2)^4}  f_{1/2}(T, m_{1/2})\right).
\end{align}
Here the factors $f_i(T, m_i) \simeq 1$ and $(m_i/T)^{3/2} e^{-m_i/T}$ for $m \gg T$ and $m \ll T$ respectively, account for the number density of the external squarks ($i=0$) and gluinos ($i=1/2$) in the process.  In the above formula, the first term arises from the production of an on-shell scalar, which subsequently oscillates and becomes a fermion via scattering.  The second term arises from an off-shell scalar, with, e.g., a quark-gluino initial state.
In the scalar mass basis, flavor changing is induced by the off-diagonal Yukawa coupling with a rate
\begin{align}
\label{eq:scalarbasis}
    \Gamma_{13,m_0} \sim {\rm min}(1,\tilde{\theta}_{13}^2) \times \alpha_s  y_b^2 T \left( f_0(T, m_0) + \tan^{-2}\beta \right),~~ \tilde{\theta}_{13} = \frac{\delta_{13} m_0^2}{y_b^2 T^2 + \tilde{m}_{33}^2-\tilde{m}_{11}^2 } .
\end{align}
Here, $y_{b}$ is the MSSM bottom quark Yukawa coupling, which is enhanced compared to the SM value by $\tan\beta$ and $\tilde{\theta}_{13}$ is the angle that diagonalizes the scalar mass matrix.  The first term in parentheses corresponds to interactions with the heavy Higgs multiplet, which we assume has a mass $m_{0}$; the second term comes from interactions with the SM Higgs boson.  For $T > m_0/y_b$, the denominator in the expression for $\tilde{\theta}_{13}$ is dominated by the thermal mass. For $T< m_0/y_b$, it may be dominated by the zero-temperature mass difference. 
We consider two extreme cases, a large tree-level mass splitting $\tilde{m}_{33}^2-\tilde{m}_{11}^2 \sim m_0^2$ and a small mass splitting generated by quantum corrections $\sim y_b^2 m_0^2$. The chiral symmetry breaking rate by the mixing of the right-handed up-type squarks or the left-handed squarks can be estimated in a similar way;
$y_b$ is replaced with $y_t$ and ${\rm tan}\beta$ is replaced with 1.
For simplicity, we assume that all squarks have the same mixing $\delta_{13}$.
Note that all chiral symmetries (in both the up and down sectors) must be broken. 
The overall suppression by the Yukawa coupling in Eq.~(\ref{eq:scalarbasis}) favors mixing in the down sector being the bottleneck, 
but the suppression of $\tilde{\theta}_{13}$ by the thermal mass $y_{t,b}^2 T^2$ can more than compensate, and the bottleneck can be in the up sector. Similarly, $\tilde{\theta}_{13}$ may be suppressed by the top Yukawa in the degenerate case where the zero temperature mass splitting is controlled by quantum corrections $\sim y_{t,b}^2 m_{0}^2$.
The washout rate of the axion rotation when the squark mixing is the bottleneck is 
\begin{align}
\label{eq:washoutMixing}
    \gamma_{13}\sim {\rm min}(\Gamma_{13,y},\Gamma_{13,m_0}) \times \frac{T^2}{S^2}.
\end{align}
For temperatures above the superpartner masses, checking for washout requires comparing this expression, along with Eqs.~(\ref{eq:washoutSS}), (\ref{eq:washoutYukawa}), and (\ref{eq:washoutGluino}), to the Hubble scale.  Washout only occurs if there is an epoch when all rates simultaneously exceed the Hubble scale. Well below the superpartner masses, the effective theory is that of the SM, and the washout rate is given by that in the SM with the QCD axion, and only Eqs.~(\ref{eq:washoutSS}) and (\ref{eq:washoutYukawa}) apply.

\begin{figure}
\includegraphics[width=0.495\linewidth]{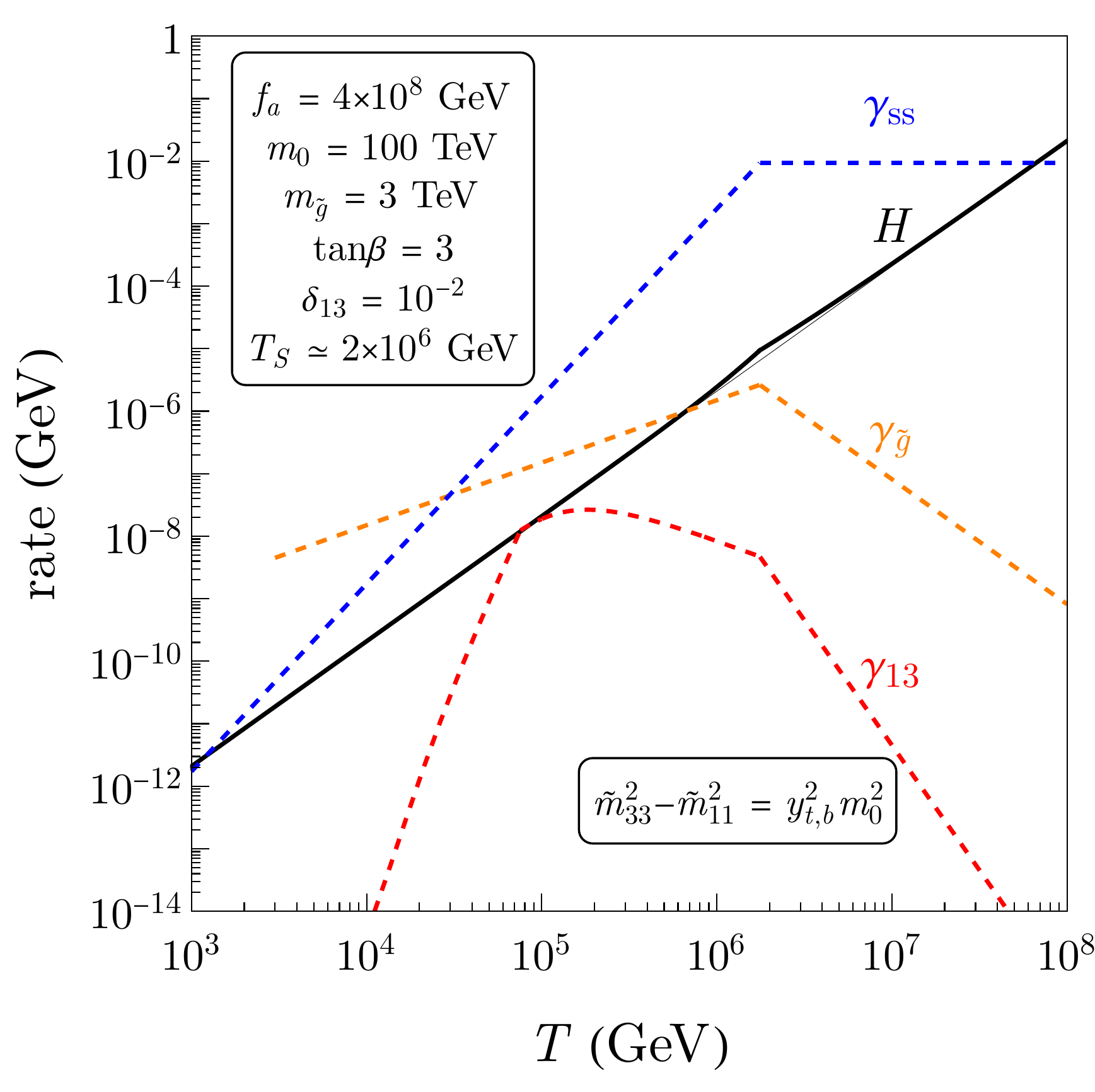}
\includegraphics[width=0.495\linewidth]{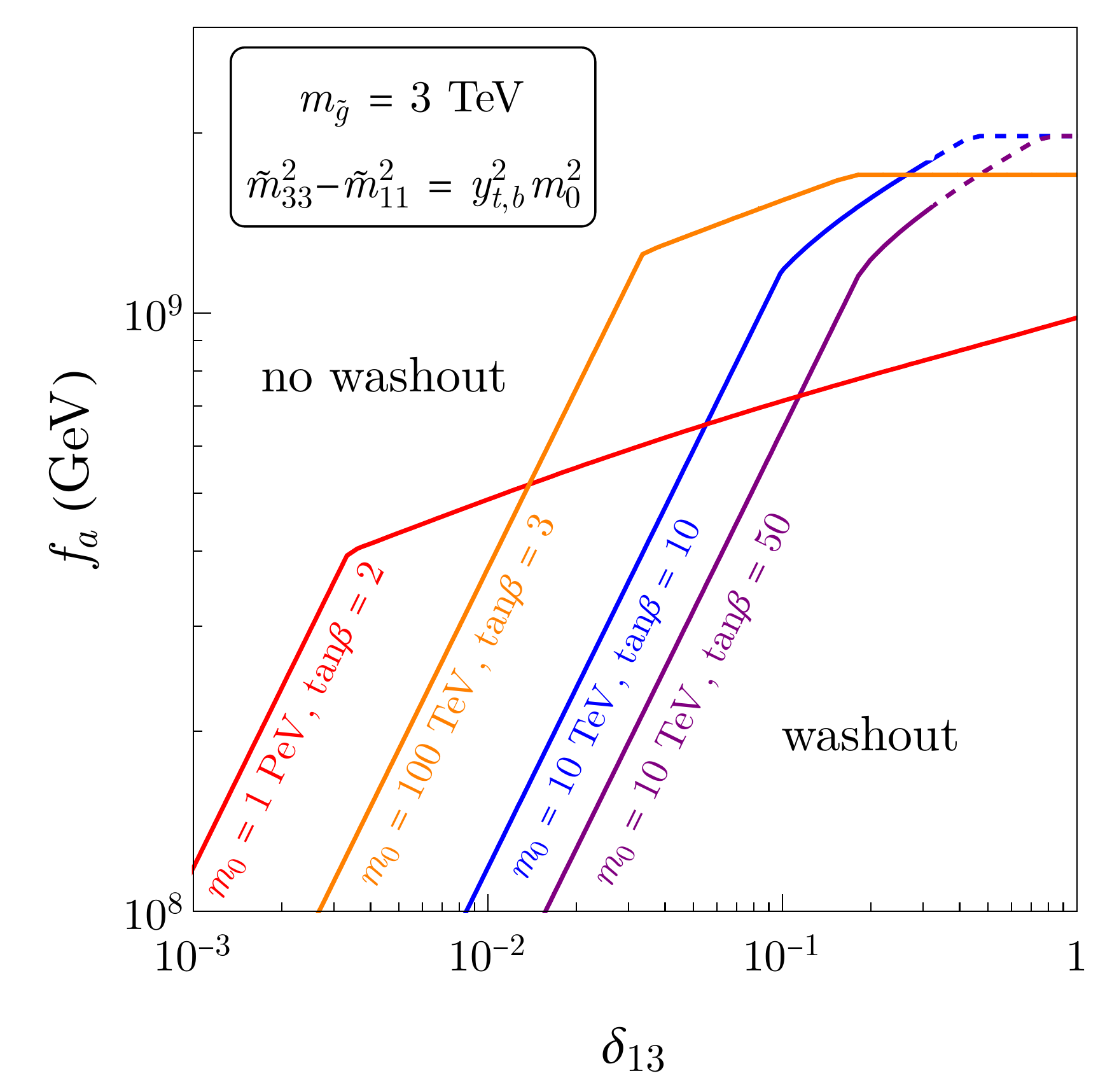}
\includegraphics[width=0.495\linewidth]{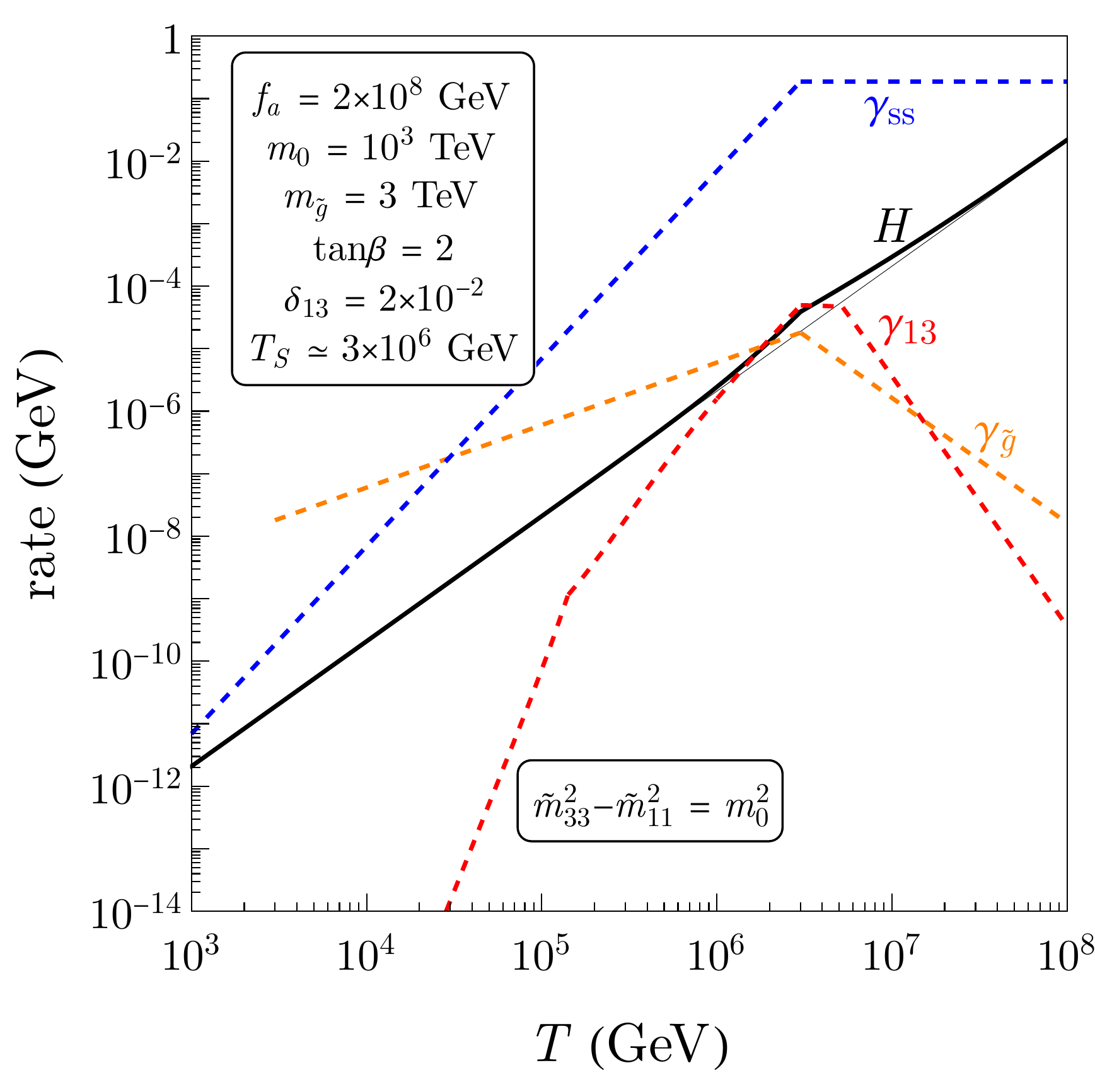}
\includegraphics[width=0.495\linewidth]{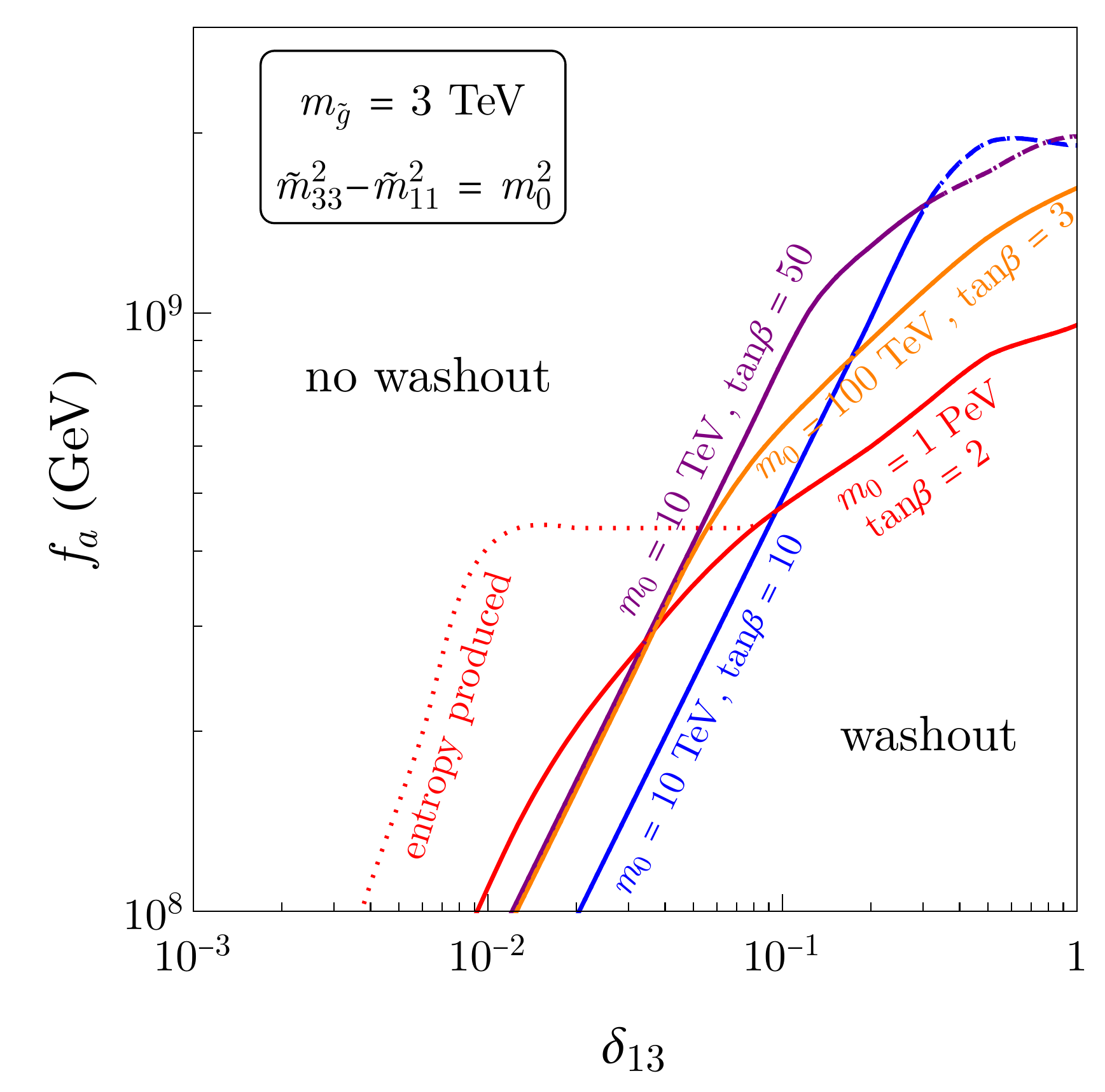}
\caption{
Left: Hubble and the washout interaction rates as a function of temperature for benchmark points where squark masses are nearly degenerate (top), or have $\mathcal{O}(1)$ splitting (bottom).
Right: Minimum values of $f_a$ required for different choices of $m_0$ and $\tan{\beta}$
to prevent washout for degenerate scalars (top) and ${\mathcal O}(1)$ splittings (bottom) as a function of squark mixing $\delta_{13}$, which we take as universal for left-handed and right-handed up and down squarks. Dashed segments indicate $\delta_{13}$ that induce too-large FCNC in the $B$ meson system. For $m_0 = 1$ PeV (bottom right panel), below the dotted line, partial washout of the axion rotation produces entropy; the estimation of the baryon asymmetry as well as the washout constraint will be modified.  In right panels, the assumption that the KMM provides the dark matter fixes $T_{S}$.}
\label{fig:washout}
\end{figure}

In the left panels of Fig.~\ref{fig:washout}, we show the Hubble rate (black line) and the individual washout rates (colored curves) for benchmarks set of parameters, chosen so that the KMM explains axion dark matter. A nearly degenerate case, where the scalar mass splitting is of the size expected to be induced by renormalization group evolution, $\tilde{m}^2_{33} - \tilde{m}^2_{11}  \sim y_{t,b}^2 m_0^2$, is shown in the top panels, and the case with generic splittings, $\tilde{m}^2_{33} - \tilde{m}^2_{11}  \sim m_0^2$, in the bottom panels.
Since all these rates depend on the saxion field value $S$, in assessing whether washout occurs, it is important to consider how $S$ changes as a function of temperature. At temperatures above $T_S$, the temperature at which the saxion settles to its minimum, see Eq.~(\ref{eq:TS_KMM}), $S \propto T^{3/2}$ and all the above rates are IR-dominated (i.e., they increase in relation to Hubble, which scales as $T^2$ in radiation domination).

Below $T_S$, $S\simeq f_a$ becomes constant; this is the origin of the kinks in the washout rates at this temperature.
The strong sphaleron washout rate in Eq.~(\ref{eq:washoutSS}) is UV-dominated. The gluino washout is IR-dominated. The washout by squark mixing in Eq.~(\ref{eq:washoutMixing}) becomes UV-dominated once the temperature is below the scalar mass $m_0$.

The axion rotation may dominate the energy density of the universe.  In this case there exists a matter-dominated era followed by a kination-dominated era. The deviation from radiation domination results in a kink in the black line in the left panels. A thin gray line with simple scaling $H \propto T^2$ is shown to guide the eye. The presence of this kination era  does not ultimately affect whether washout occurs in the degenerate case (top panels).  In the case of the non-degenerate sfermions (bottom panels), the kination era can indeed impact washout. We make additional comments on this below.

In the right panels of Fig.~\ref{fig:washout}, we show the lower bounds on $f_a$ as a function of $\delta_{13}$ for different specified values of $m_0$ and $\tan\beta$
with a fixed gluino mass $m_{\tilde{g}}$ of 3 TeV. For each $(f_a,m_0)$, $T_S$ is chosen so that the KMM explains axion dark matter. Again, we show this for both a degenerate case (top) and non-degenerate case (bottom).  If at least one process is out of equilibrium, washout is avoided.  With a sufficiently small $\delta_{13}$, the suppressed rate $\gamma_{13}$ in Eq.~(\ref{eq:washoutMixing}) is never larger than the Hubble scale; this is responsible for the portion of the colored curves with the steepest slope.
In this case, the lower bound of $f_a$ is proportional to $\tilde{\theta}_{13} \tan\beta$. 
On the other hand, for sufficiently large values of $f_a$, the rates $\gamma_{\rm ss}$ and $\gamma_{\tilde{g}}$ in Eqs.~(\ref{eq:washoutSS}) and (\ref{eq:washoutGluino}), respectively, are never simultaneously larger than the Hubble expansion rate, which prevents washout and removes the bound on $\tilde{\theta}_{13} \tan\beta$. This is the case for the horizontal segments of the blue and purple curves in the upper right panel, and in this limit the lower bound on $f_a$ scales with $m_{\tilde{g}}^{1/2}$. For the intermediate values of $f_a$ and $\tilde{\theta}_{13} \tan\beta$ (negatively-sloped segments in blue and purple), the strong sphaleron process goes out of equilibrium right before the squark mixing washout comes into equilibrium. Lastly, the horizontal and negatively-sloped segments of the orange curve are when the gluino washout comes into equilibrium after the squark mixing process goes out of equilibrium. The curve turns horizontal at large $\delta_{13}$ because $\gamma_{13}$ transitions from being set by $\Gamma_{13,y}$ in Eq.~(\ref{eq:Yukawabasis}) to $\Gamma_{13,m_0}$ in Eq.~(\ref{eq:scalarbasis}) (for down-type squarks) with $\tilde{\theta}_{13}$ already saturated to unity.

Squark mixing is also constrained by flavor physics.
Using the formulae in~\cite{Gabbiani:1996hi} and the bounds on the $B$ meson mixing derived in~\cite{Charles:2020dfl}, we find $\delta_{13} < 0.3$ for $m_0 = 10$ TeV, while it can be $\mathcal{O}(1)$ for $m_0 \gtrsim 20$ TeV. Here we take the limit $m_{\tilde{g}}\ll m_0$. The excluded values of $\delta_{13}$ are indicated by the dashing of the lines in the right panels of Fig.~\ref{fig:washout}.

In a similar manner, we show regions where washout occurs for the non-degenerate sfermions in the lower right panel. The washout effect is generically suppressed compared to the degenerate case (upper right panel) due to the following two factors. First, the mixing rate is suppressed by the smaller $\tilde\theta_{13}$ in Eq.~(\ref{eq:scalarbasis}).  Consequently, the steepest segments of the curves shift to larger $\delta_{13}$. Second, unlike the degenerate case, whether the bottleneck washout rate exceeds the Hubble expansion rate is sometimes determined during the epoch of matter and kination domination (around the kink in the black line in the left panels). The enhanced Hubble expansion rate relative to a radiation-dominated case weakens the bound on $f_a$.  An example of this phenomenon is shown in the bottom left panel; the gluino rate $\gamma_{\tilde{g}}$ and squark mixing rate $\gamma_{13}$, the bottleneck processes in this case, peak during this era.

Finally, since washout is not always negligible during kination/matter domination, we need to check whether entropy is produced from the rotation during such an era---even if washout is not complete. There can exist a time where the radiation created from the washout processes exceeds that of the pre-existing radiation, $\rho_{\rm rot} \times \min \left(1, \gamma_{\rm WO} / H \right) > \rho_{\rm rad}$, where $\gamma_{\rm WO} = \min \left( \gamma_{\rm ss}, \gamma_{\tilde{g}}, \gamma_{13} \right)$. Therefore, even though washout is incomplete, our assumption of no entropy production from washout may be violated, and the corresponding derivation of the baryon asymmetry would need to be revisited.  This occurs in the region below the dotted red curve in the bottom right panel. As the washout rates depend on temperature, the extra radiation accelerates the washout, and the true bound on $\delta_{13}$ to avoid complete washout of the rotation would lie somewhere between the two red curves.

To summarize, in both the degenerate and non-degenerate cases, the washout of the axion rotation is avoided for $f_a\gsim 10^9$ GeV even if the squark mixing is $\mathcal{O}(1)$. For $f_a \lsim 10^9$ GeV, avoiding washout puts an upper bound on squark mixing stronger than the one from flavor physics.  We emphasize that this washout analysis is not peculiar to RPV axiogenesis but applies to any cosmological scenario that includes axion rotations, e.g., the KMM, in a supersymmetric setup.

\nocite{apsrev41Control}
\bibliographystyle{apsrev4-1}
\bibliography{RPV}

\end{document}